\documentclass[9pt,shortpaper,twoside,web]{dentalfeature}
\usepackage{dentalfeature}
\usepackage{cite}
\usepackage{amsmath,amssymb,amsfonts}
\usepackage{algorithmic}
\usepackage{graphicx}
\usepackage{textcomp}
\usepackage{times}
\usepackage{epsfig}
\usepackage{subfigure}
\usepackage{verbatim}
\usepackage{color}
\usepackage{ragged2e}
\usepackage{multicol}
\usepackage{gensymb}
\usepackage{tabularx}
\usepackage{overpic}
\usepackage[switch]{lineno}
\newcommand{\etal}{\textit{et al.}}
\def\BibTeX{{\rm B\kern-.05em{\sc i\kern-.025em b}\kern-.08em
    T\kern-.1667em\lower.7ex\hbox{E}\kern-.125emX}}
\begin{document}
\title{Dense Representative Tooth Landmark/axis Detection Network on 3D Model}
\author{Guangshun Wei, Zhiming Cui, Jie Zhu, Lei Yang, Yuanfeng Zhou, Pradeep Singh, Min Gu, Wenping Wang \IEEEmembership{Fellow, IEEE} 
\thanks{Manuscript received November, 2021; 
%This work was supported in part by the U.S. Department of Commerce under Grant BS123456. 
}
\thanks{Guangshun Wei and Yuanfeng Zhou are with the School of Software, Shandong University, Jinan 250101, P.R.China. (e-mail: guangshunwei@gmail.com; yfzhou@sdu.edu.cn). }
\thanks{Zhiming Cui and Lei Yang are with the Department of Computer Science, and University of Hong Kong, Hong Kong SAR, China. (e-mail: cuizm.neu.edu@gmail.com; yanglei.dalian@gmail.com). }
\thanks{Wenping Wang, Texas A\&M University, and University of Hong Kong, Hong Kong SAR, China. (e-mail: wenping@cs.hku.hk). }
\thanks{Jie Zhu is with the Department of Computer Science, Nanjing University, Nanjing 210093, P.R.China. (e-mail: magickuang@126.com). }
\thanks{Pradeep Singh and Min Gu are with the Faculty of Dentistry, The University of Hong Kong, Hong Kong SAR, China. (e-mail: pradeepal1928@gmail.com; drgumin@hku.hk). }
}

\maketitle

\begin{abstract}
Artificial intelligence (AI) technology is increasingly used for digital orthodontics, but one of the challenges is to automatically and accurately detect tooth landmarks and axes. This is partly because of sophisticated geometric definitions of them, and partly due to large variations among individual tooth and across different types of tooth. As such, we propose a deep learning approach with a labeled dataset by professional dentists to the tooth landmark/axis detection on tooth model that are crucial for orthodontic treatments. Our method can extract not only tooth landmarks in the form of point (e.g. cusps), but also axes that measure the tooth angulation and inclination.
The proposed network takes as input a 3D tooth model and predicts various types of the tooth landmarks and axes. Specifically, we encode the landmarks and axes as dense fields defined on the surface of the tooth model. This design choice and a set of added components make the proposed network more suitable for extracting sparse landmarks from a given 3D tooth model. Extensive evaluation of the proposed method was conducted on a set of dental models prepared by experienced dentists. Results show that our method can produce tooth landmarks with high accuracy. Our method was examined and justified via comparison with the state-of-the-art methods as well as the ablation studies.
\end{abstract}


\section{Introduction}
\label{sec:introduction}

\begin{figure}[th]
\centering
\subfigure[Misaligned teeth]{
\includegraphics[width=3.5cm]{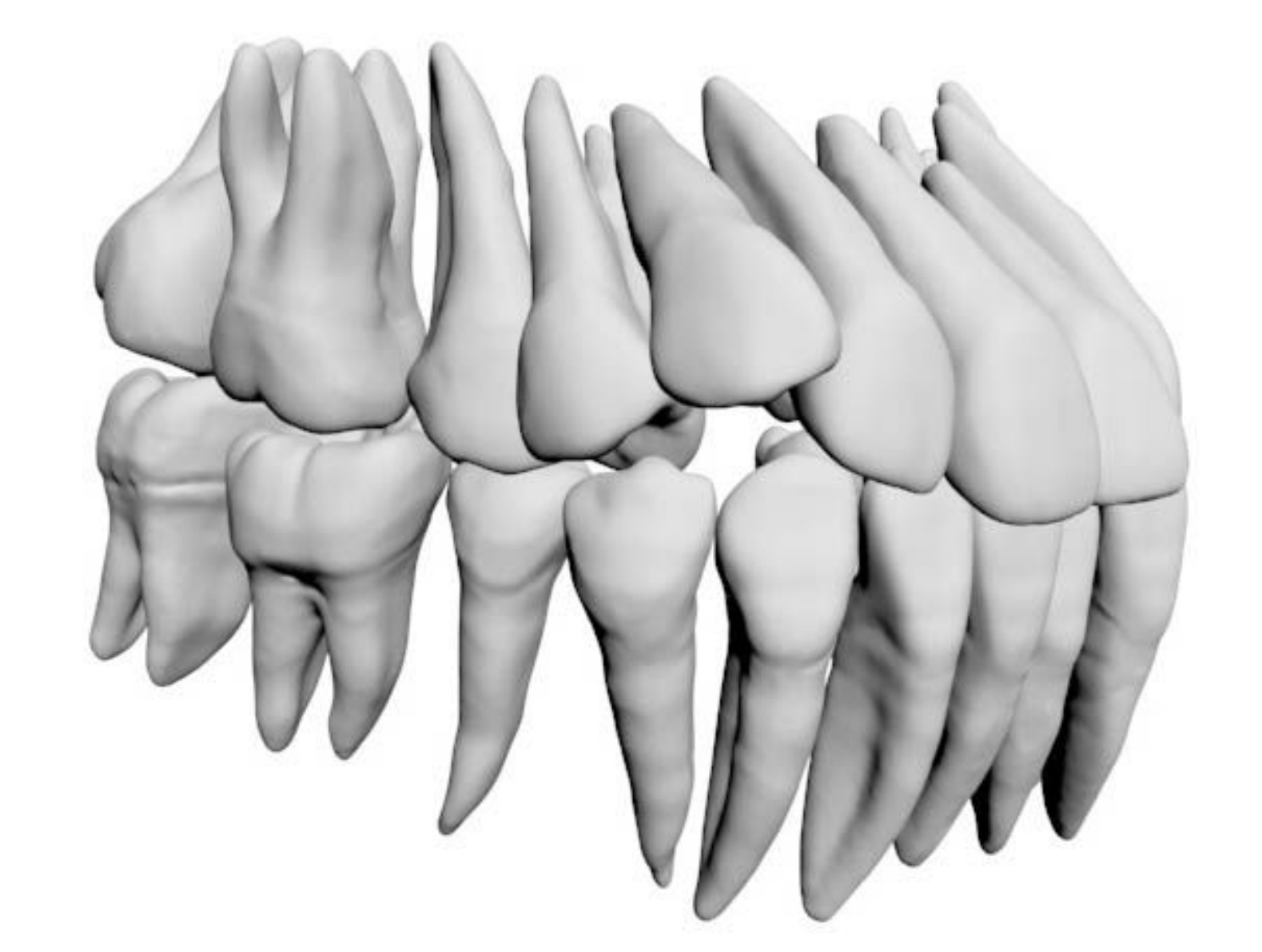}
}
\subfigure[The result of orthodontic treatment]{
\includegraphics[width=3.5cm]{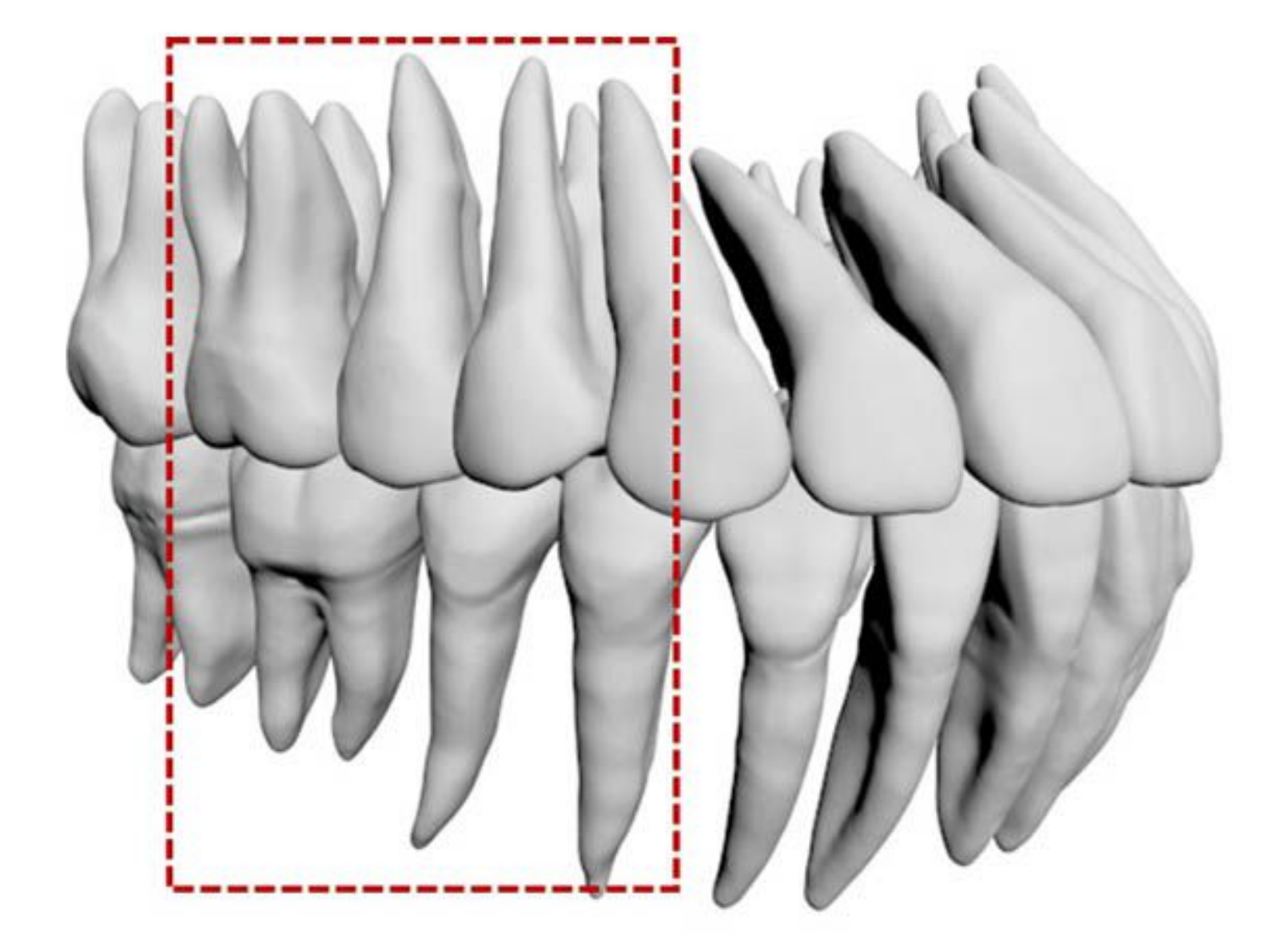}
}
\subfigure[OC \& FA]{
\includegraphics[width=1.75cm]{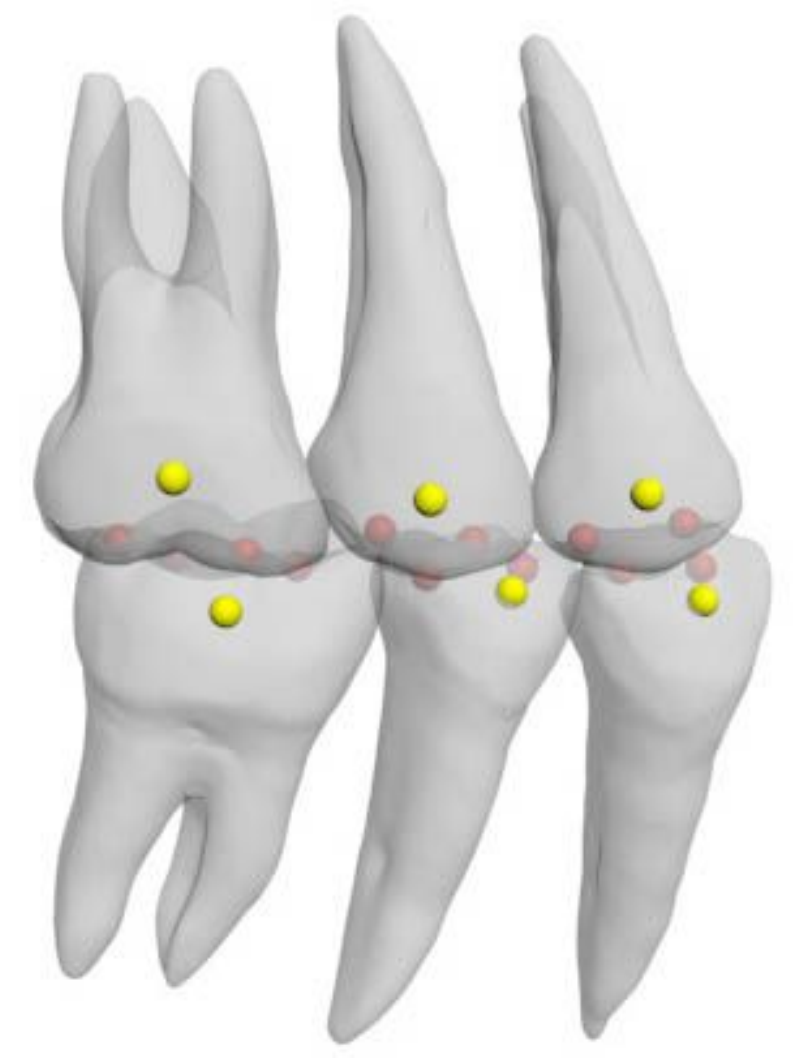}
}
\subfigure[CO \& CU]{
\label{OCCO}
\includegraphics[width=1.75cm]{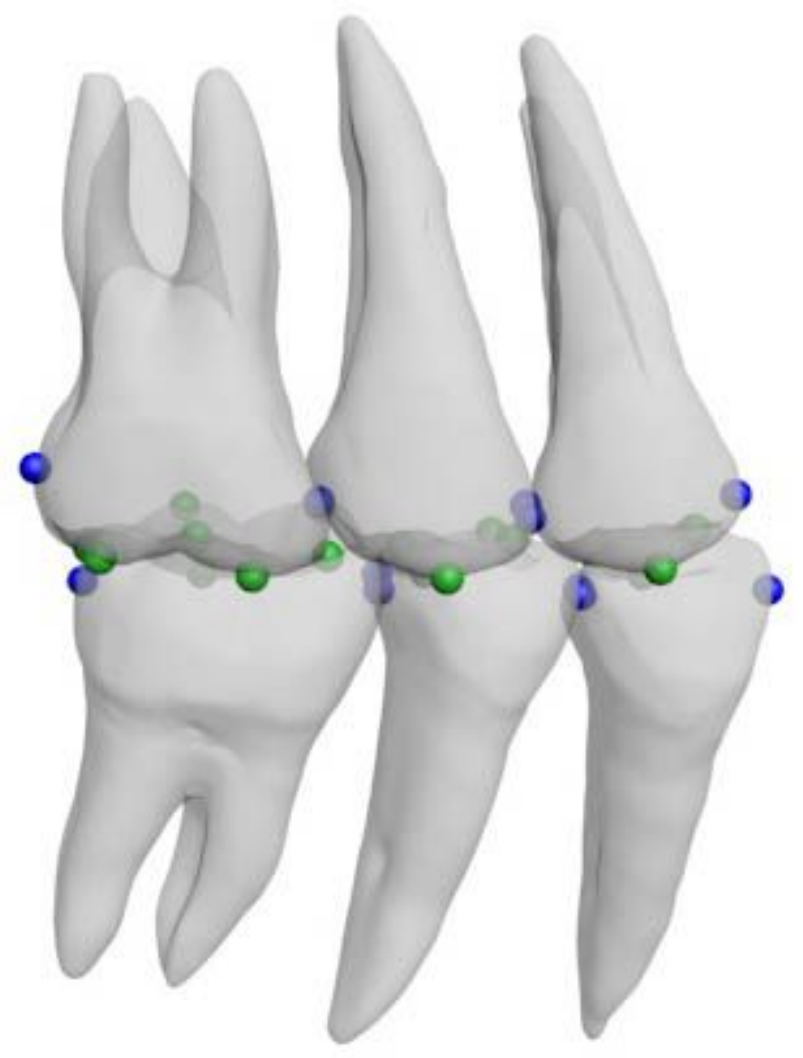}
}
\subfigure[BA \& LA]{
\includegraphics[width=1.75cm]{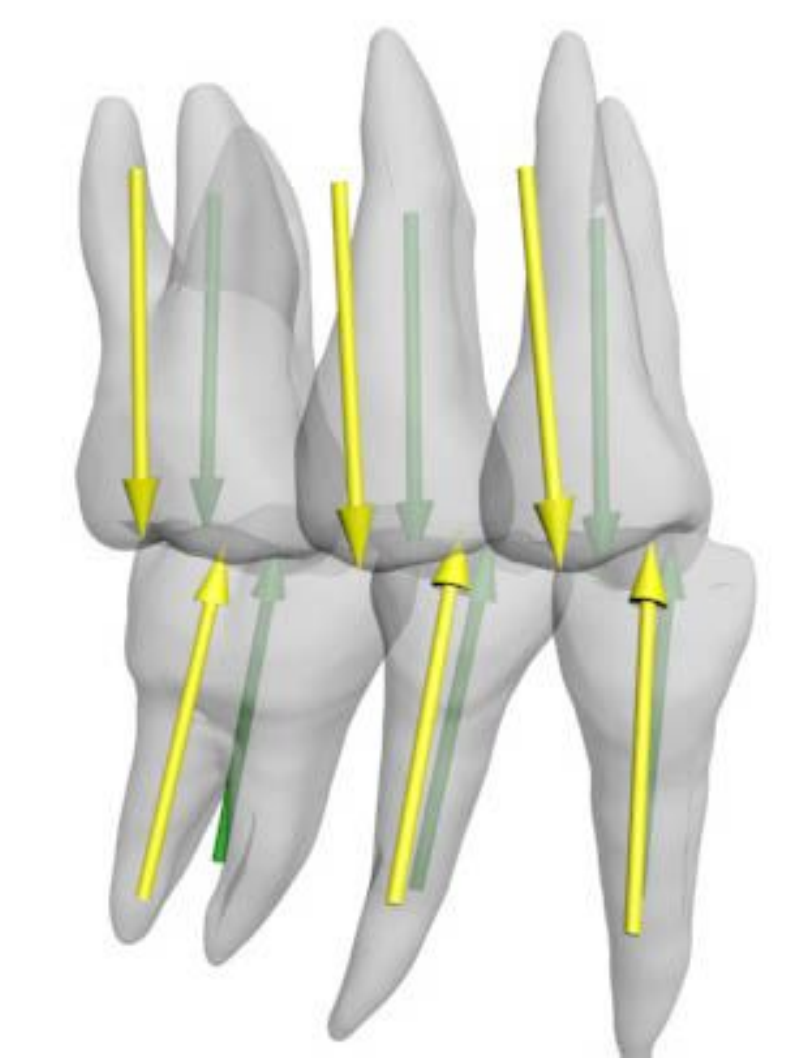}
}
\subfigure[MA \& DA]{
\includegraphics[width=1.75cm]{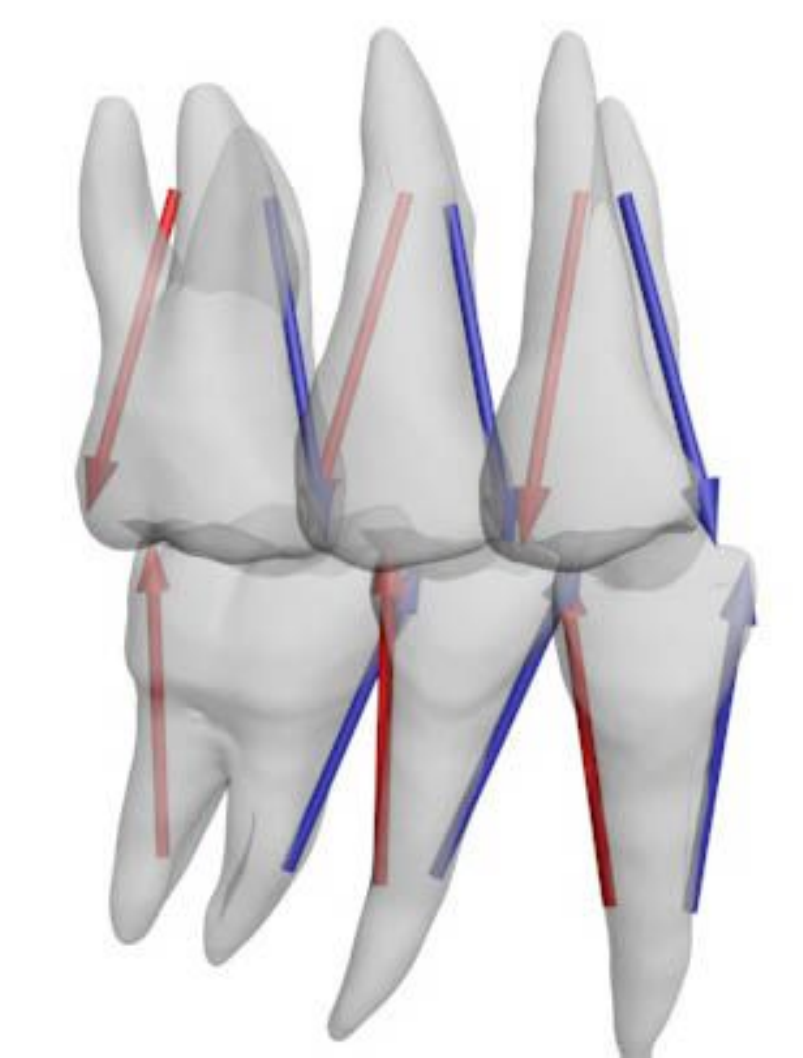}
}
\caption{Using tooth landmarks and axes defined in \textit{Andrews' Six Keys} to characterize the deviation of current tooth configuration from normal occlusion. The red, yellow, blue, and green dots represent occlusal point (OC), facial axis point (FA), contact point (CO), and cusp point (CU), respectively. The green, yellow, blue, and red arrows represent lingual axis (LA), buccal surface axis (BA),  mesial axis (MA), and distal axis (DA), respectively. (a) shows the state of misaligned teeth. (b) shows the desired arrangement after orthodontic treatment based on tooth landmarks and axes. (c)-(f) show the role and significance of dental landmarks and axes on orthodontics.}
\label{fig:dental_features}
\vspace{-0.3cm}
\end{figure}

\begin{table}[h]
    \centering
    \caption{Description of tooth landmarks and axes~\cite{None2017The}.}
    \begin{tabular}{lp{0.6\columnwidth}lllllll}
    \hline
     
       Contact point (CO)   &  The point of contact between the proximal surfaces of the two adjacent teeth. \\
      \hline
      Cusp point (CU)  & Extremity of the cone-shaped protuberance on the crown of a tooth.\\
      \hline
      Facial axis point (FA) & Centre/mid point of the facial surface (approximating either the lips or the cheek) of a tooth.\\
      \hline
      Occlusal point (OC)     & The point of normal contact between opposing teeth, when the maxilla and mandible are brought together during the habitual occlusion/act of closure. \\  
      \hline
      Buccal surface axis (BA)  & A tangent is drawn along the slope of the tooth surface that is adjacent to the lip or cheek.\\
      \hline
      Lingual axis (LA)   & A tangent is drawn along the tooth surface facing the tongue and parallel to the long axis of the tooth.\\
      \hline
      Mesial axis (MA)  & A tangent is drawn along the proximal surface that is faced toward the central line of dental arch.\\
      \hline
      Distal axis (DA)   & A tangent is drawn along the proximal surface that is faced away from the center line of dental arch.\\
    \hline
    \end{tabular}
    \label{tab:definition_of_dental_features}
\end{table}

\begin{figure*}[h]
%\vspace{-3cm}
\begin{center}
\includegraphics[width=6.5in]{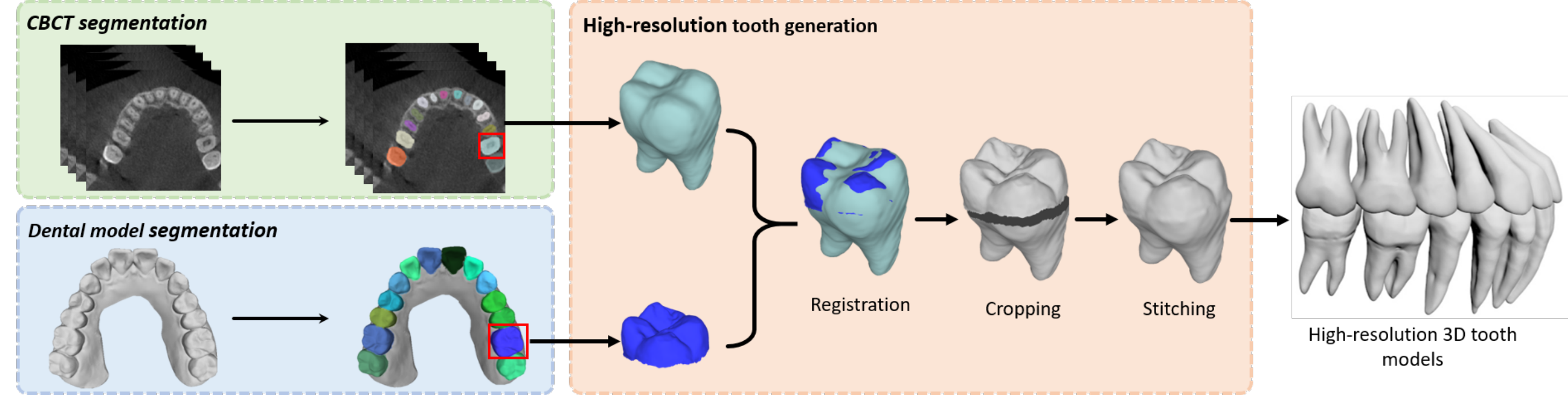}
\end{center}
   \caption{Details of data processing. The strategy of registration, cropping and stitching is used to generate a 3D tooth model with high-precision crown and root information based on CBCT data and oral scan data. }
\label{fig:data_process}
\vspace{-0.3cm}
\end{figure*}

Malocclusion refers to the misalignment of teeth when the jaws are closed. In severe cases, it could lead to discomfort, or even speech or breathing problems. In order to treat this disorder, the primary goal of the dental treatment is to align the teeth based on certain empirical rules/conditions, such as \emph{Andrews Six Keys}~\cite{andrews1972six}, that are considered to be central for achieving a normalized occlusion and fulfillment of these, results in successful orthodontic treatment.
In a nutshell, such sufficient conditions can be effectively represented as a set of desired relationships between the \textit{pre-defined tooth landmarks and/or axes} that should be satisfied in the normal occlusion (see Fig.~\ref{fig:dental_features}). For example, the contact points of each tooth at the normal occlusion can be used to align adjacent teeth and itself at the current configuration (see Fig.~\ref{OCCO}), while the relationship between the pair of cusp and occlusal points and the pair of their corresponding points on the opposite jaw signifies the cusp-groove, cusp-fossa or cusp-embrasure relationship of the upper and lower jaws.

With automatized technologies gaining popularity by the use of digital diagnostic aids and customized appliances especially the clear aligners~\cite{AIorthodontics,AIrole},  there has been a heightened demand for Artificial Intelligence (AI) techniques to optimize contemporary orthodontic treatment planning. Many existing methods~\cite{5482178,choi2012validity,cheng2015personalized,Piece} have been proposed to enhance the automatic level of the orthodontic treatment and save the time of clinicians. However, the detection of the aforementioned tooth landmarks and axes that are crucial for defining the normal occlusion is largely left to experienced dentists and performed manually. 
 Only a few studies~(e.g.,~\cite{kumar2012automatic,sun2020automatic}) explored the possibility of automatic detection of dental landmarks for teeth alignment.
To further boost the automation level of tooth alignment, this paper aims to provide an automatic, data-driven approach to extract, from different types of teeth, both the landmarks and the axes that are used in \textit{Andrews' Six Keys}, as exemplified in Fig.~\ref{fig:dental_features} and listed in Table~\ref{tab:definition_of_dental_features}.

A major challenge is that most tooth landmarks and axes do not correspond to any sharp geometry on the given tooth model. For example, the contact points lie in smooth regions of the tooth surface (see Fig.~\ref{fig:dental_features}), and all tooth axes are defined with both global and local geometry of the tooth. Thus, Kumar~\etal~\cite{kumar2012automatic}, a geometry-based method, can only localize the cusp point, a sharp geometry on the tooth crown, by combining the surface curvature and height value.

With the recent advancement of deep learning, some learning based methods, e.g.,~\cite{2020An}, have been proposed in this specific task of tooth alignment. Unfortunately, these methods are still limited to detect landmarks with distinctive geometry (e.g., cusp point), which are insufficient to define a normal occlusion and thus far from practical for the orthodontic treatment. 
Another stream of learning based methods, including PointNet++~\cite{qi2017pointnet++} and SpiderCNN~\cite{xu2018spidercnn}, are designed to perform recognition or classification for general geometric processing tasks. However, they are less accurate when applied to our specific task of detecting tooth landmarks and axes that are sparse signals (landmarks or axes) of the given mesh model (see Table \ref{tab:compareaxistableothernetwork}).

To address the above issues, we propose a novel neural network for reliably detecting tooth landmarks and axes on a 3D tooth model. To this end, we built a dataset covering different teeth with their landmarks and axes for orthodontic treatment by professional dentists.
Given a tooth landmark on the 3D tooth surface, we first convert it to a geodesic distance field defined on the surface of the tooth model. Thus, localization of the tooth landmark can be converted to first predicting a distance field on the given tooth surface and then detecting the peak of the distance field. 
Similarly, for a given tooth axis, we represent it with a bundle of projection vectors, each of which projects the associated point on the surface to the tooth axes. These dense representations are collectively termed as point-wise field coding in the paper.
Considering that the landmarks and axes of a tooth depend on both local and global geometry of a tooth, we utilize a multi-scale mechanism to learn latent feature representations by setting different scale radii on the input point cloud and aggregating them to extract the hierarchical latent features.
In addition, we employ an enhancement module that concatenates the spatial coordinates, the normal of the input points, and the extracted latent features to further improve the prediction results.

Our contributions are summarized as follows:

\begin{itemize}
\item We present a novel framework that robustly detects tooth landmarks and axes from a 3D tooth model. 
%This is the first work to address this challenging and meaningful research work.
\item We propose a new mechanism to encode sparse geometric entities (i.e. points and axes) into dense representations, and further adopt a multi-scale and enhancement module to learn reliable latent features based on this dense representations.
\item We conducted extensive experiments and comparisons based on a set of dental models prepared by experienced dentists. The average errors for tooth landmark and axis detection by our method are 0.37 $mm$ of the tooth size and 3.33\degree, respectively. We also demonstrate the advantages of our method quantitatively and qualitatively via comparative studies with state-of-the-art methods and ablation studies.
\end{itemize}

\section{Related work}

In recent years, many digital methods and tools have been proposed to facilitate digital dentistry, ranging from tooth segmentation~\cite{gan2015toward,cui2019toothnet}, dental diagnosis~\cite{raith2017artificial}, restoration~\cite{beckett2005preservation}, cephalometric landmark detection~\cite{chen2019cephalometric}, and orthodontic treatment process~\cite{2020An,wei2020tanet}.

\begin{figure*}[h]
%\vspace{-3cm}
\begin{center}
\includegraphics[width=6.5in]{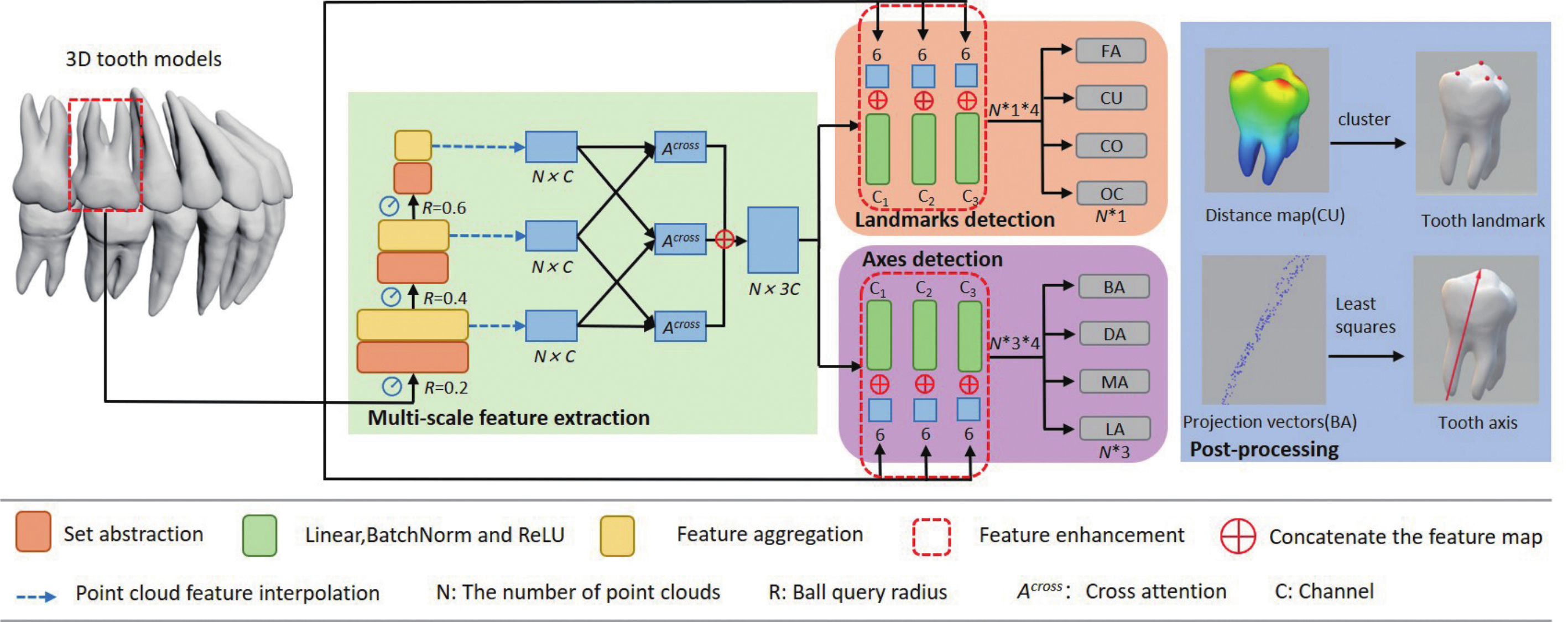}
\vspace{-0.3cm}
\end{center}
   \caption{The pipeline of the proposed method. Given the point cloud of the 3D tooth model based on data processing as input, we initially extract the feature vectors through the multi-scale feature extraction. The extracted feature map is then simultaneously delivered to the two branch sub-networks. Then the distance field and projection vectors are predicted by the two sub-networks. Finally, tooth landmarks and axes are obtained through the feature enhancement module and post-processing steps, respectively.  }
\label{fig:pipeline}
%\vspace{-0.2cm}
\end{figure*}

\noindent\textbf{Deep learning on Point Cloud.} Most existing works convert point clouds to images or volumetric forms~\cite{Wu20153D,Maturana2015VoxNet} before feature learning. However, these CNN based methods still require quantization of point clouds with voxel resolution and high requirements for hardware devices. Recently, PointNet~\cite{Charles2017PointNet} is proposed to use the DNN with raw point clouds as input, for the purposes of classification and segmentation, and provides interesting theoretical insight into processing point clouds. Subsequently, PointNet++~\cite{qi2017pointnet++} is proposed as an improved version of pointnet, using a hierarchical feature learning and multi-scale point cloud convolution operations. Subsequently, other models of point cloud-based deep neural networks have also been proposed~\cite{xu2018spidercnn,wu2019pointconv,li2018pointcnn,wang2019dynamic}. Furthermore, with the development of computer graphics in deep learning~\cite{mitra2018creativeai,MitraRKGKRY18}, many methods tried to connect the point set into a graph and extract the latent features. ECC~\cite{simonovsky2017dynamic} used dynamic edge condition filters, where the convolution kernel is based on the edge generation inside the point cloud. GAC~\cite{wang2019graph} proposed a novel graph attention convolution, whose kernel can be dynamically converted into a specific shape to adapt to the structure of the model. Recently, transformer network models PCT~\cite{guo2020pct} and PT~\cite{zhao2020point} based on point cloud were proposed, and they performed well in point cloud classification and segmentation tasks. Furthermore, many fields based on point cloud learning have received extensive attention, such as normal estimation~\cite{GuerreroKOM18,LiuFMLXP19}, point cloud registration~\cite{abs210304256}, point cloud super-resolution~\cite{ShanWCSE20}, and point cloud denoising~\cite{RakotosaonaBGMO20,ChenWSXW20} etc.

\noindent\textbf{Keypoint detection.}
There are many methods for keypoint detection, including traditional methods and more recent data-driven methods. 
Most of the traditional methods~\cite{Maes2010Feature,Li2013Curvature,Godil2011Salient,Sipiran2010A} use local geometric information, such as curvature or surface normals, for keypoint detection. This line of works assumes that the keypoints are related to or even defined by the geometric change, and thus can work well to detect sharp features in mechanical parts. However, these methods cannot robustly handle complex 3D shapes (e.g., tooth) with large shape variations.
An alternative is the data-driven approach.
For 2D images, the key point detection (e.g., face~\cite{Maes2010Feature} and hand~\cite{simon2017hand}) usually transforms sparse landmarks into heatmaps and set them as the target of supervised learning. 
For 3D models, the supervised algorithm in~\cite{shu2018detecting} predicts the probability distribution defined on the surface. However, these methods usually suffer from many artifacts, due to the dependence on local salient features and local geometric changes, especially when the tooth characteristics are not prominent.

\noindent\textbf{Tooth landmark and axis detection.} In recent years, digital orthodontics has become a research focus. However, given the constraints of complex alignment rules and the patient’s individualized functional aesthetics, malocclusion~\cite{jafri2020digital} poses a challenge to orthodontists. The timely and effective treatment of mal-aligned teeth depends on the quality of the landmarks and axes detection. The conception of \textit{Andrews' Six Keys}~\cite{andrews1972six} was the earliest proposed tooth-related feature. Then, with the advent of digital technologies and computer assisted tooth movement, clinicians have been able to perform comprehensive and deliberate evaluation resulting in enhanced esthetics and improved treatment outcomes~\cite{lu2009improving,finelle2017digital}. However, these methods were manual or semi-automatic. None of the existing methods~\cite{Zhang2011Kinematics,Busch2006Concept} can fully detect all the landmarks and axes due to the challenges in data acquisition and the inconspicuous characteristics of the tooth surface. In this paper, we propose a fully automated and efficient method to detect landmarks and axes based on deep learning.

\section{Methods}

In this section, we first describe data processing and the network architecture on point cloud learning, and then introduce the two branches for detecting the tooth landmarks and axes, respectively. Finally, we present the implementation details of the network. Fig.~\ref{fig:pipeline} shows the pipeline of our method.

\subsection{Data processing}\label{sec:data_preparation}
To obtain a 3D tooth model for our specific task, we leverage both the intra-oral scanned data and the CBCT data to reconstruct 3D models of individual teeth.
CBCT data provide comprehensive 3D information of a tooth (including crown and root) and is suitable for 3D tooth model acquisition. However, the tooth model reconstructed from CBCT data loses the fine geometric details of the crown. On the other hand, the intra-oral scanned data can capture the high-precision geometric details of the tooth crown. For tooth landmarks detection, most dental landmarks are located in the crown region. For tooth axis detection, it pays more attention to the overall shape of the tooth. Therefore, we use the strategy of data processing to obtain the final high-precision 3D tooth model.

Specifically, as shown in Fig.~\ref{fig:data_process}, given the paired CBCT image and dental model scanned from a patient in clinics, we first adopt ToothNet \cite{cui2019toothnet} and TSegNet \cite{cui2021tsegnet} to faithfully segment tooth and tooth crown from these two modality data. 
Then, a rigid ICP registration is applied to align the crown parts between the two surfaces, and the tooth crown obtained from CBCT is cropped and removed.
Finally, with the aligned tooth root and crown, we obtained a 3D tooth model with high-precision crown and root information by using screened poisson method~\cite{kazhdan2013screened} to complete and reconstruct the tooth. Note that for some failed cases in the data pre-processing process, we will remove them manually to ensure the accuracy of the subsequent tooth landmark and axis detection process.

\begin{figure*}
    \centering
    \includegraphics[width=6.5in]{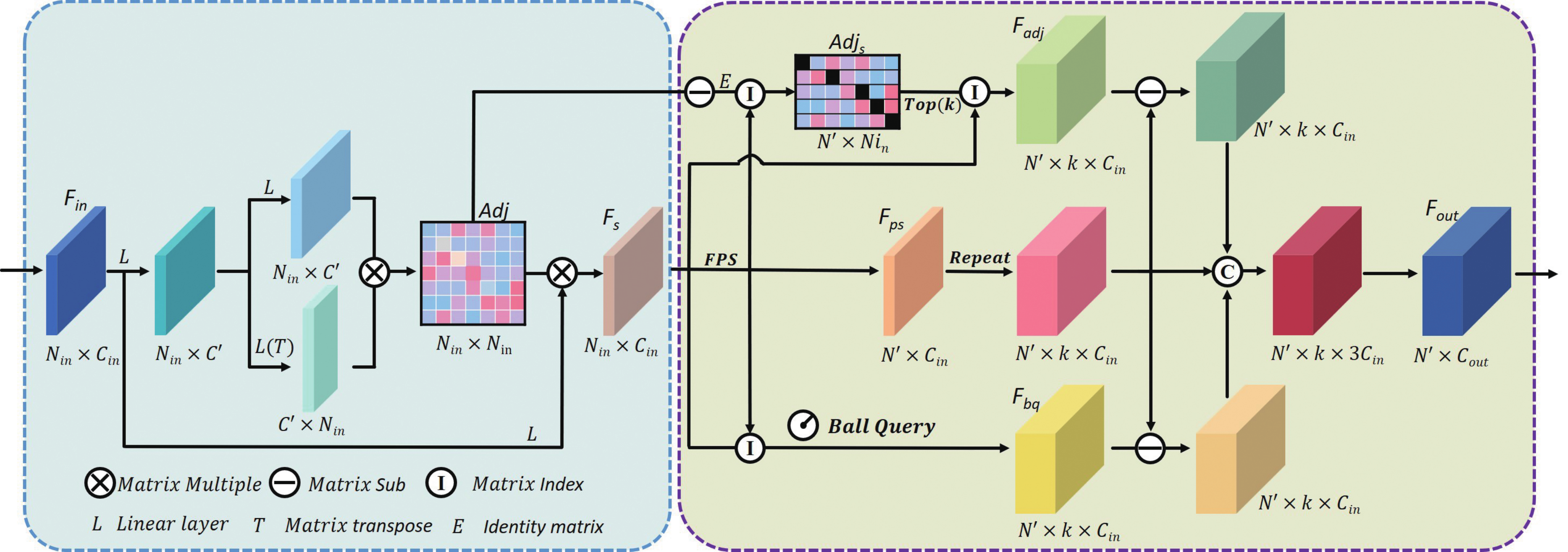}
    \caption{Point cloud feature extraction and feature aggregation module. FSP is the farthest point  sampling  strategy.}
    \label{fig:feat_agg} 
\end{figure*}

\subsection{Network}
As shown in Fig.~\ref{fig:pipeline}, our network consists of three major components to predict the dense representations of the tooth landmarks and axes, including a multi-scale latent feature extraction module and two sub-networks for tooth axis and landmark detection, respectively. We will elaborate on each module in detail.

\subsubsection{multi-scale latent feature extraction module} 
Although our method is designed for two different tasks (i.e., tooth landmark and axis detection), they are intrinsically related from a geometric perspective. Thus, given a point cloud of a 3D tooth model acquired as described in Sec.~\ref{sec:data_preparation}, we first extract common point-wise features by a multi-scale latent feature extraction module. 
Importantly, as the localization of the dental landmarks and axes relies on the crown geometry and the holistic tooth shape, both local and global information are needed. For example, the cusp point mainly depends on local crown geometry whereas the tooth axes mainly depends on holistic tooth shape information. 

As shown in Fig.~\ref{fig:pipeline}, firstly, we adopt a multi-scale feature extraction scheme to capture local information with different radii. During this process, we propose a novel \textbf{feature aggregation module}, which aggregates the features of point clouds in both local and non-local spaces. %Different from the other methods, our strategy can effectively aggregate local and non-local features. 
As shown in Fig.~\ref{fig:feat_agg}, given the input feature map $F_{in}$, through the feature extraction layer $L$, an adjacent matrix $Adj$ is obtained by the $Softmax \left( L(F_{in}) \times L(F_{in})^T \right)$. Each element in $Adj$ indicates the similarity of different point-wise features in $F_{in}$.
The new feature map $F_{s}$, capturing non-local shape information, is then calculated by $Adj \times L(F_{in})$. 
Subsequently, a sub-sampled point cloud $p_s$ with point-wise features are obtained by the farthest point sampling (FPS) strategy, and the corresponding adjacency matrix is calculated by $ Adj_s= Index(Adj - E)$ where $E$ is a identity matrix.
For each point on $p_s$, we select the $Top (k)$ points based on each row of the adjacency matrix $Adj_{s}$ to obtain aggregated features $F_{adj}$.
Also, to capture the feature in local space, for each point on $p_s$, we adopt Ball Query to search nearest $k$ points within the radius $r$ to obtain aggregated features $F_{bq}$.
At last, the original feature $F_{ps}$, non-local feature $F_{adj}$, and local feature $F_{bq}$ of the sub-sampled point cloud $p_s$ are combined together for final prediction.

\begin{equation}
	\begin{array}{lr}
	F_{cat}=concat\left( F_{ps}, F_{adj}-F_{ps},F_{bq}-F_{ps}\right), & \\[1mm]
	F_{out}=MP(MLP(F_{cat})), & \\[1mm]
	\end{array}
\end{equation}
where $MP$ is the max-pooling operator, and $MLP$ is the multi-layer perceptron.

After each feature aggregation module, the hierarchical feature with the same number of input point clouds are obtained through interpolation. To further capture shape and context information of the point cloud, \textbf{cross attention} layer is applied at different scales to extract multi-scale features, which is defined as follows:
\begin{equation}
A^{cross}\left( x,y,y\right)=\alpha \left( \frac{ QK^{T}}{\sqrt{d_k}}\right)V,	
	\left\{  
	\begin{array}{lr}
	Q=Conv1D(x), & \\[1mm]
	K=Conv1D(y), & \\[1mm]
	V=Conv1D(y), & \\[1mm]
	\end{array}
	\right.
\end{equation}
where $x, y$ refers to the hierarchical feature and $d_k$ is the  $K$ of dimension to increase stability. 
Finally, we concatenate the extracted all features together, and then feed the decoded point-wise features into the following two sub-networks to predict the distance fields and projection vector fields, respectively.

\subsubsection{Feature enhancement module}
With the common point-wise features, two task-specific modules are adopted to predict the tooth axes and landmarks. As shown in Fig.~\ref{fig:pipeline}, since the detection of tooth landmarks and axes is sensitive to spatial coordinates and normals, we propose a feature enhancement module to concatenate the input point coordinates and normals with the extracted latent features. Specifically, green boxes indicate the latent features extracted by the multi-scale feature extraction module, and blue boxes represent the coordinate information and normal information of the input points. We then combine these two information as inputs to LBR modules (Linear, BatchNorm and ReLU) for finally predicting the distance fields of landmarks and projection vector fields of axes.

\subsection{Tooth landmarks detection}
\subsubsection{Coding landmarks with point-wise distance fields} %(\textbf{difficult,our idea}) 
Different from other geometric landmarks detection problems, most of our landmarks are defined without salient local geometric features, as shown in Fig.~\ref{fig:dental_features}. For example, the facial axis point lies in a smooth region of the tooth buccal side, where no sharp feature exists. Under this circumstance, a straightforward regression method with target coordinates provides insufficient supervision and usually leads to ambiguous predictions. Inspired by face landmark detection in the image domain, in which researchers used a heatmap regression strategy to reduce such ambiguity~\cite{dong2018style,zhang2018combining}. We extend this idea to the 3D domain by encoding landmarks to each point of a tooth model to achieve better performance.

\begin{figure}[h]
\centering
\subfigure{
\includegraphics[width=1.75cm]{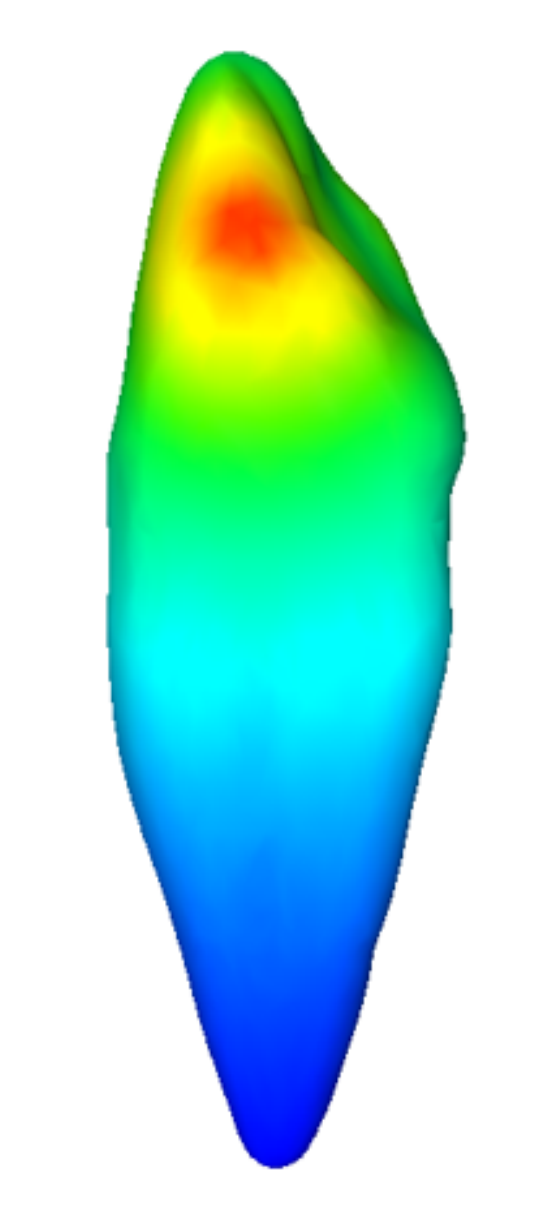}
}
\subfigure{
\includegraphics[width=1.75cm]{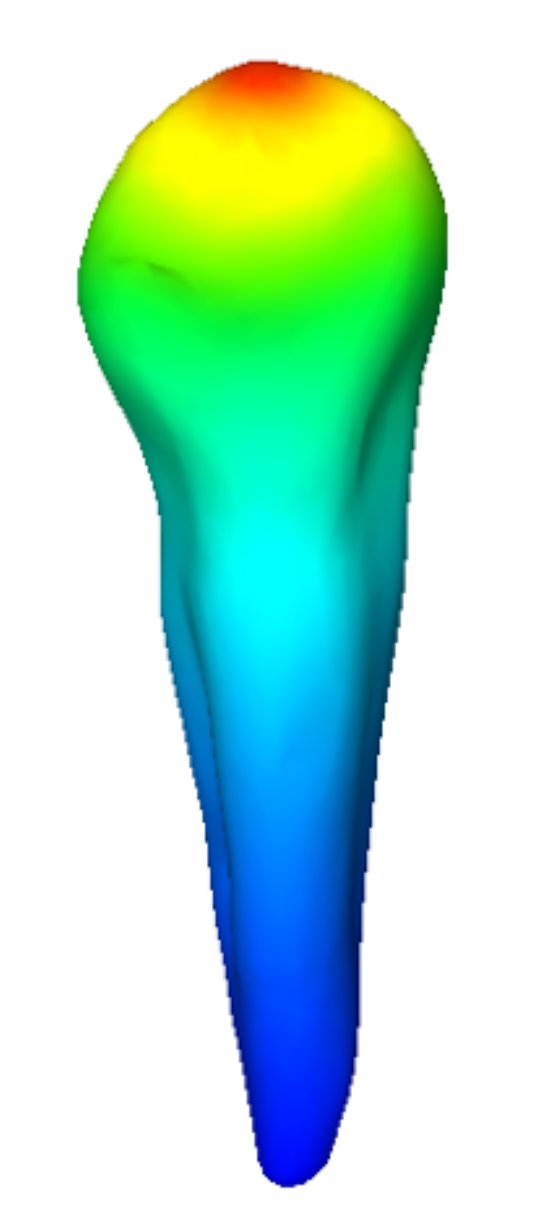}
}
\subfigure{
\includegraphics[width=1.75cm]{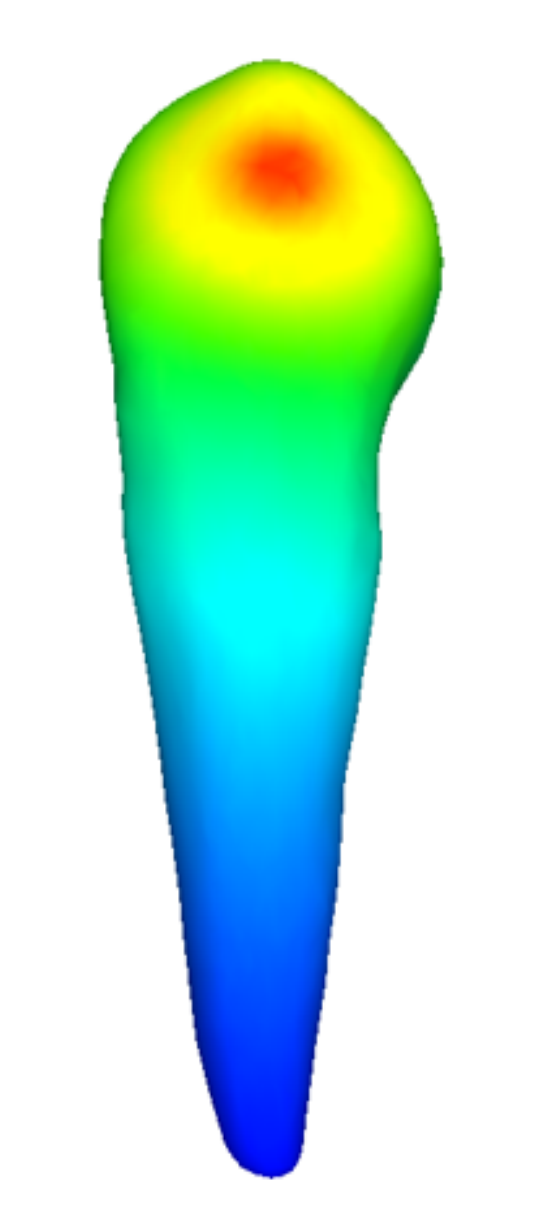}
}
\subfigure{
\includegraphics[width=1.75cm]{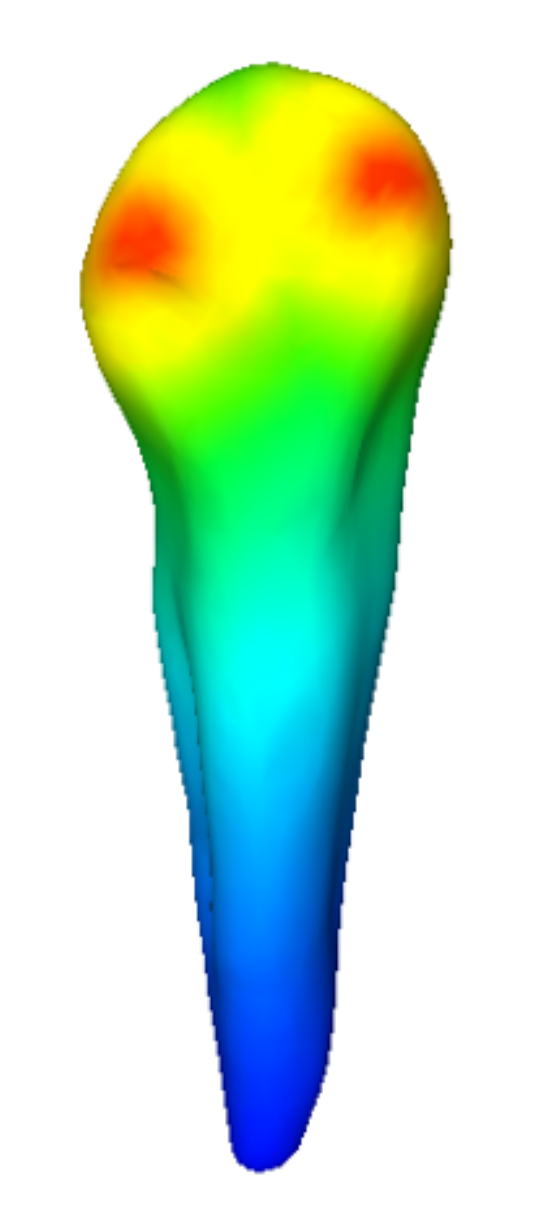}
}
\caption{Distance fields corresponding to different landmarks, where in turn represents a contact point, a cusp point, a facial axis point and an occlusal point.}
\label{fig:distacemap}
%\vspace{-0.5cm}
\end{figure}

For the landmark detection, we define a distance field $D$ as the ground truth by calculating the geodesic distance~\cite{xin2009improving} from each point $p_{i}$ to the landmark $f_{pj}$ on the 3D tooth mesh as follows:
\begin{equation}\label{equ:caldistance}
D_{i} =exp \left(  -\frac{(G(f_{pj},p_{i}))^2}{2\sigma^2}\right),
\end{equation}
where $G(\cdot)$ function aims to calculate the geodesic distance of two points on the surface of teeth, and $\sigma$ is set to 0.3 in our method. We get the final distance field $D$ by Eq.~\ref{equ:caldistance}. If there are multiple landmarks on the tooth model, we combine the calculated multiple distance fields by selecting the largest distance value for each point. As shown in Fig.~\ref{fig:distacemap}, the smaller the geodesic distance of the point is, the larger $d_{i}$ (red color) is, and vice versa.

\begin{figure}
\begin{center}
 \subfigure[]
    {
    \includegraphics[width=2.1cm]{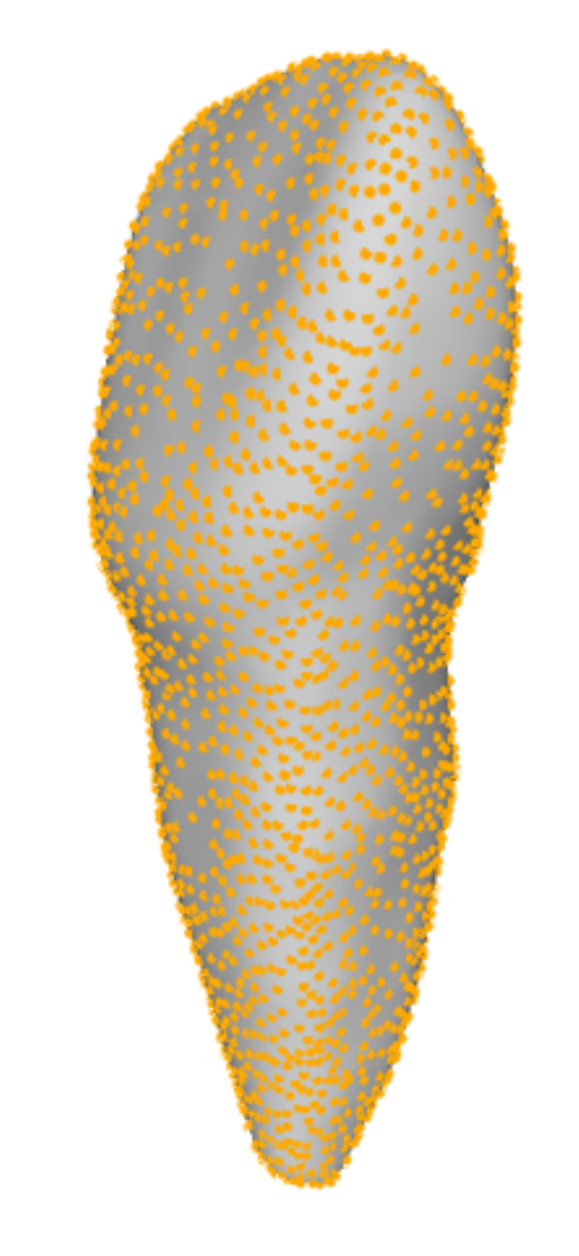}
    }
     \subfigure[]
    {
    \includegraphics[width=2.1cm]{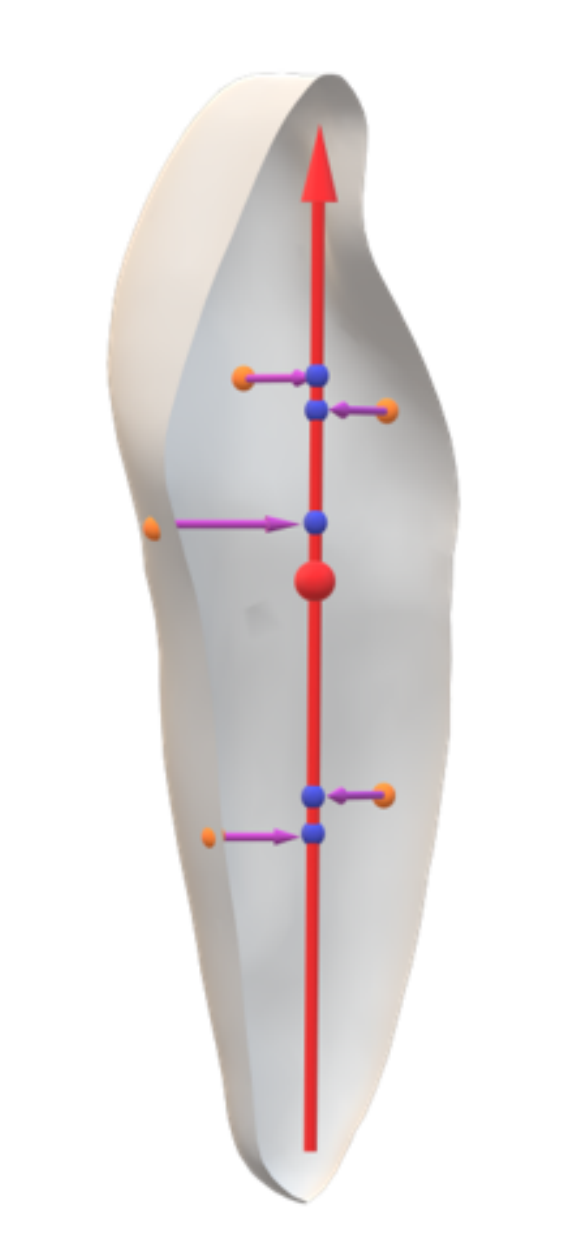}
    }
     \subfigure[]
    {
    \includegraphics[width=2.1cm]{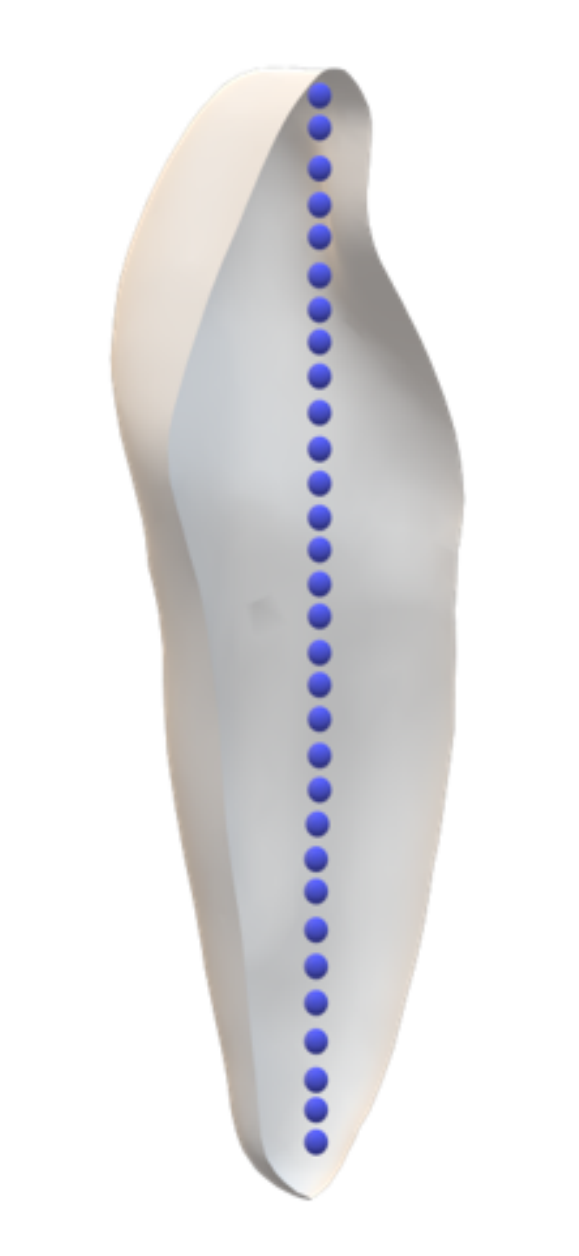}
    }
\end{center}
\caption{Point-wise field coding for tooth axes detection, where (a) is the 3D tooth model, (b) represents the process of getting the projection vectors and (c) is the final encoding features corresponding to each point. The orange dots represent the point cloud on the surface of the tooth. The red dot represents the center point of the tooth. The red arrow indicates the tooth axis that passes through the center point. The blue dots indicate the projection point at which the points on the surface of the tooth are projected perpendicularly to the tooth axes.}
\label{fig:pointcloud}
%\vspace{-0.6cm}
\end{figure}

\subsubsection{Distance field prediction}
After passing through the feature enhancement module, different distance fields on the tooth surface are outputted. We utilize the mean square error as the loss function to supervise the training process, as follows:

\begin{equation}\label{callosspoint}
\mathcal{L}_{p} =\frac{ \begin{matrix} \sum_{i=1}^N (D_{i}-D'_{i})^2 \end{matrix} }{N},
\end{equation}

\noindent where $D'_{i}$ is the predicted distance value of point $p_i$, $D_{i}$ refers to the ground truth value and $N$ is the number of points.

\subsubsection{Landmarks generation}
Given the predicted distance field of each tooth, we first obtain some partial points with relatively larger distance values. 
In other words, the distance field can be regarded as a confidence field, where the greater value of a point, the greater credibility of the landmark located on the point. 
Then we set a threshold to filter out these points with low confidence. Finally, K-Means clustering~\cite{Bahmani2012Scalable} is used according to the number of landmarks on the tooth model. The clustering centers are regarded as landmarks.

\subsection{Tooth axes detection}
Similar to the tooth landmark detection, the proper dense representation of tooth axis can also provide more abundant supervision information in the training process and robust to the outliers, compared to directly axis regression.

\subsubsection{Coding tooth axes with point-wise projection vector fields}
For tooth axes detection task, we first define the dense projection vector field of a tooth axis according to the given linear equation:
\begin{equation}
\begin{array}{l}{    
p_{center}=\frac{\begin{matrix} \sum_{i=1}^N p_{i} \end{matrix} }{N},
}\\\\
{
\mathbf{l}=p_{center}+\mathbf{n}\cdot t,
}
\end{array}    
\end{equation}
where $p_{center}$ is the center point of the tooth, $\mathbf{n}$ is the unit vector of tooth axes, and $\mathbf{l}$ is the line of the tooth axis with the modulus $t$. All points $p_{i}$ on the tooth surface are then projected perpendicularly onto this tooth axis, $\mathbf{l}$, and the corresponding project vectors are obtained as supervision signals. In other words, as shown in Fig.~\ref{fig:pointcloud}, the output of the tooth axis detection network is the projected coordinates of the point cloud.

\subsubsection{Projection vector field prediction} 
The point-wise field coding of the tooth axes is reflected in the displacement from the point on the tooth surface to its axes. Our network architecture of tooth axes detection can be seen in Fig.~\ref{fig:pipeline}.Taking the point-wise features as input, the mapping displacement relationship between the tooth surface and tooth axes is predicted. Finally, we get the different projection vectors. 
We calculate the $L_2$ distance between corresponding points of ground truth and predicted projection vectors as loss function:

\begin{equation}
    \mathcal{L}_{axis} = \frac{1}{N} \sum_{i=1}^N \| \mathbf{v}(p_i) - \mathbf{v}(p'_i) \|,
\end{equation}
\noindent where $\mathbf{v}(p_i)$ is the ground-truth projection vector at $p_i$ and $\mathbf{v}(p'_i)$ is its estimated result.

\subsubsection{Tooth axes generation}
By learning the displacement relationship from the tooth surface to the tooth axes, we get the projection vectors as shown in the branch of tooth axes detection in Fig.~\ref{fig:pipeline}. 
We then project the input point cloud with their predicted projection vectors to approximate the target tooth axes. In this paper, we use the least squares method~\cite{Sanford1980Application} to fit the line and set it as the final result of the target tooth axes.

\subsection{Implementation Details}
\subsubsection{Network details}
The input data are a set of points with normals $(N\times6)$ for a single 3D tooth model. In the multi-scale feature extraction module, we extract point cloud features by the feature aggregation and cross attention module. Then the three sets of global features from different scales are concatenated together and sent to two sub-networks. After multi-scale feature extraction, a latent feature matrix $(N \times 192)$ is obtained. Then it connects the coordinates of each point cloud $(N \times (192+6))$ and sends it to the sub-networks separately. Finally, the corresponding distance fields $(N \times 1 \times 4)$ and projection vector fields $(N \times 3 \times 4)$ are generated through the feature enhancement module $(192-64-32)$. 

\subsubsection{Training details}
We implemented the entire network in the PyTorch, using the Adam solver~\cite{kingma2014adam:} with a fixed learning rate of 0.001, batch size 32. 
We train the entire network for 300 epochs, which takes around 4 hours for training with one Nvidia GeForce RTX 3090 GPU. Note that we train an individual model for each tooth category (incisor, canine, premolar, and molar), and the total training time is about 16 hours.

\section{Experiments and Discussions}

In this section, we provide extensive ablation studies and comparisons to validate the effectiveness of our network for landmark and axis detection. 

\begin{figure}[h]
\centering
\subfigure[Incisor]{
\includegraphics[width=4cm]{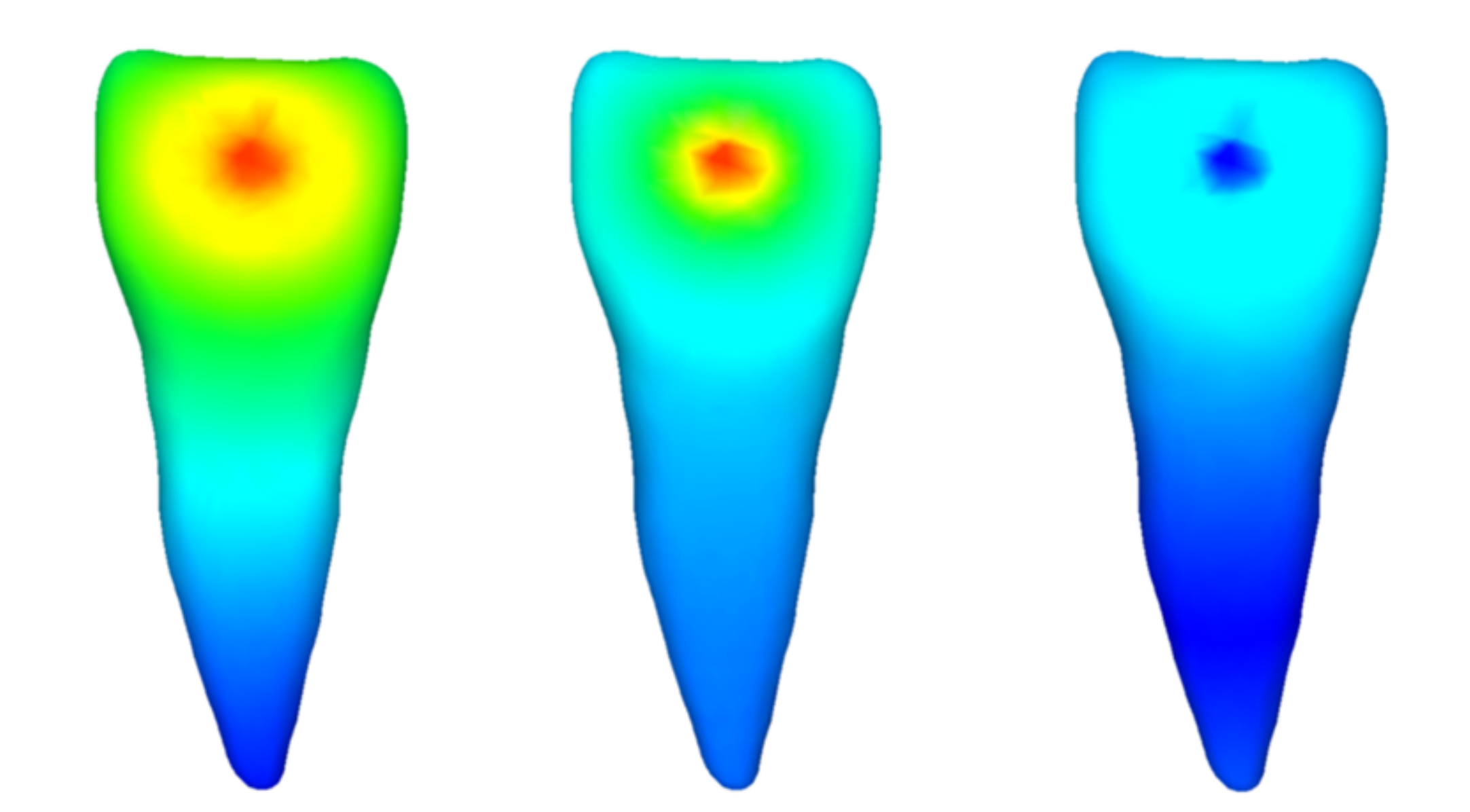}
}
\subfigure[Canine]{
\includegraphics[width=4cm]{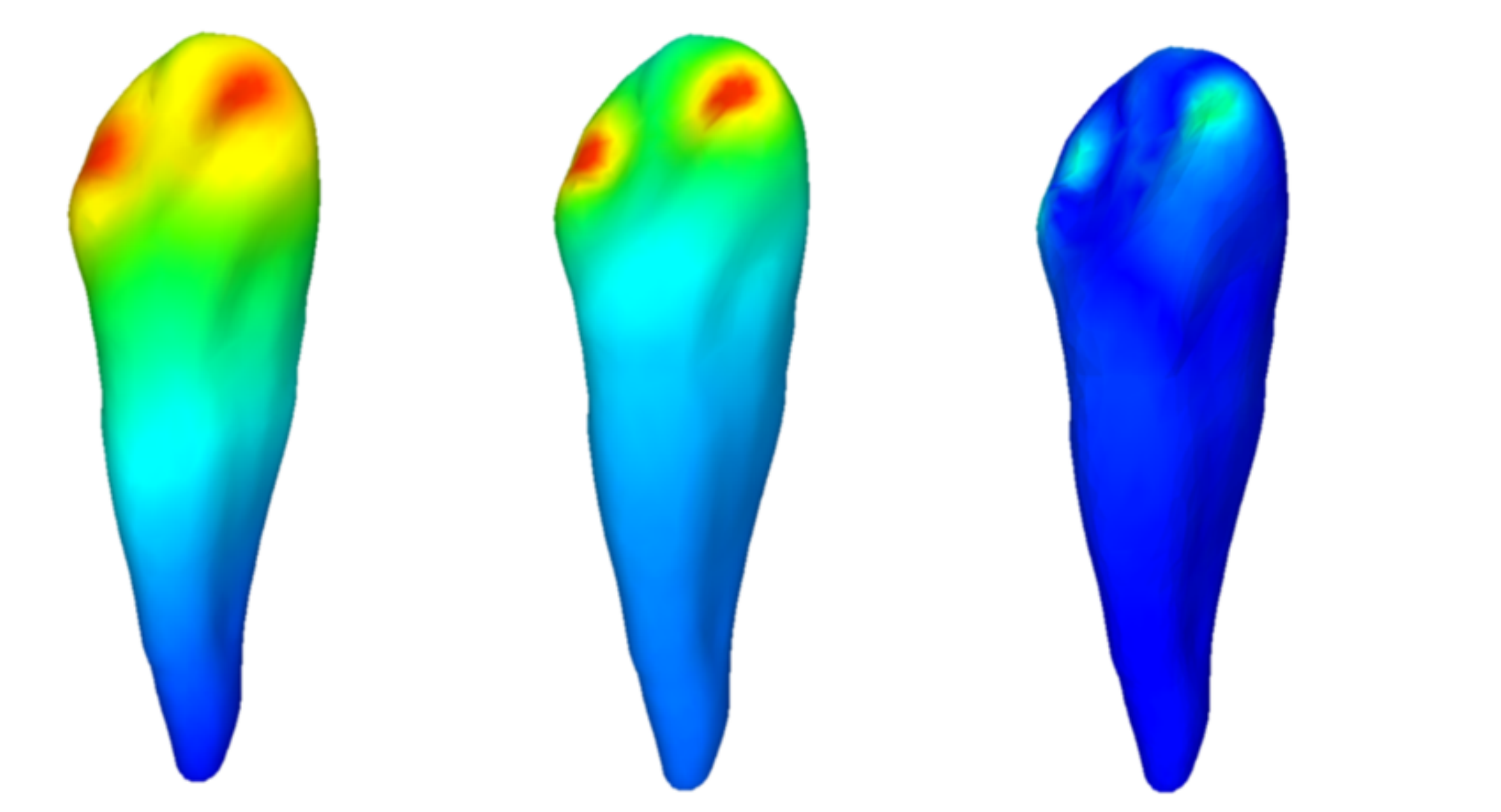}
}
\subfigure[Premolar]{
\includegraphics[width=4cm]{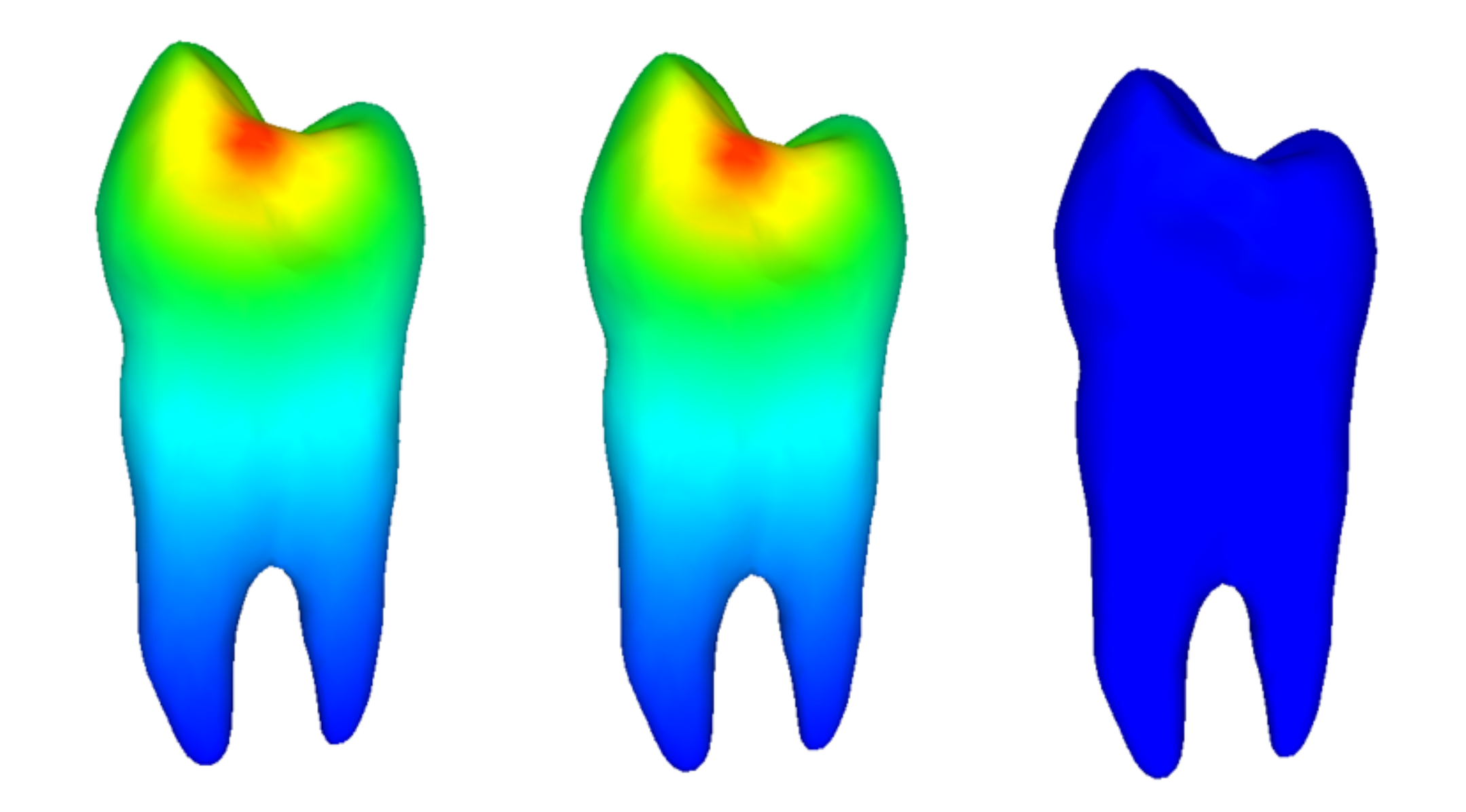}
}
\subfigure[Molar]{
\includegraphics[width=4cm]{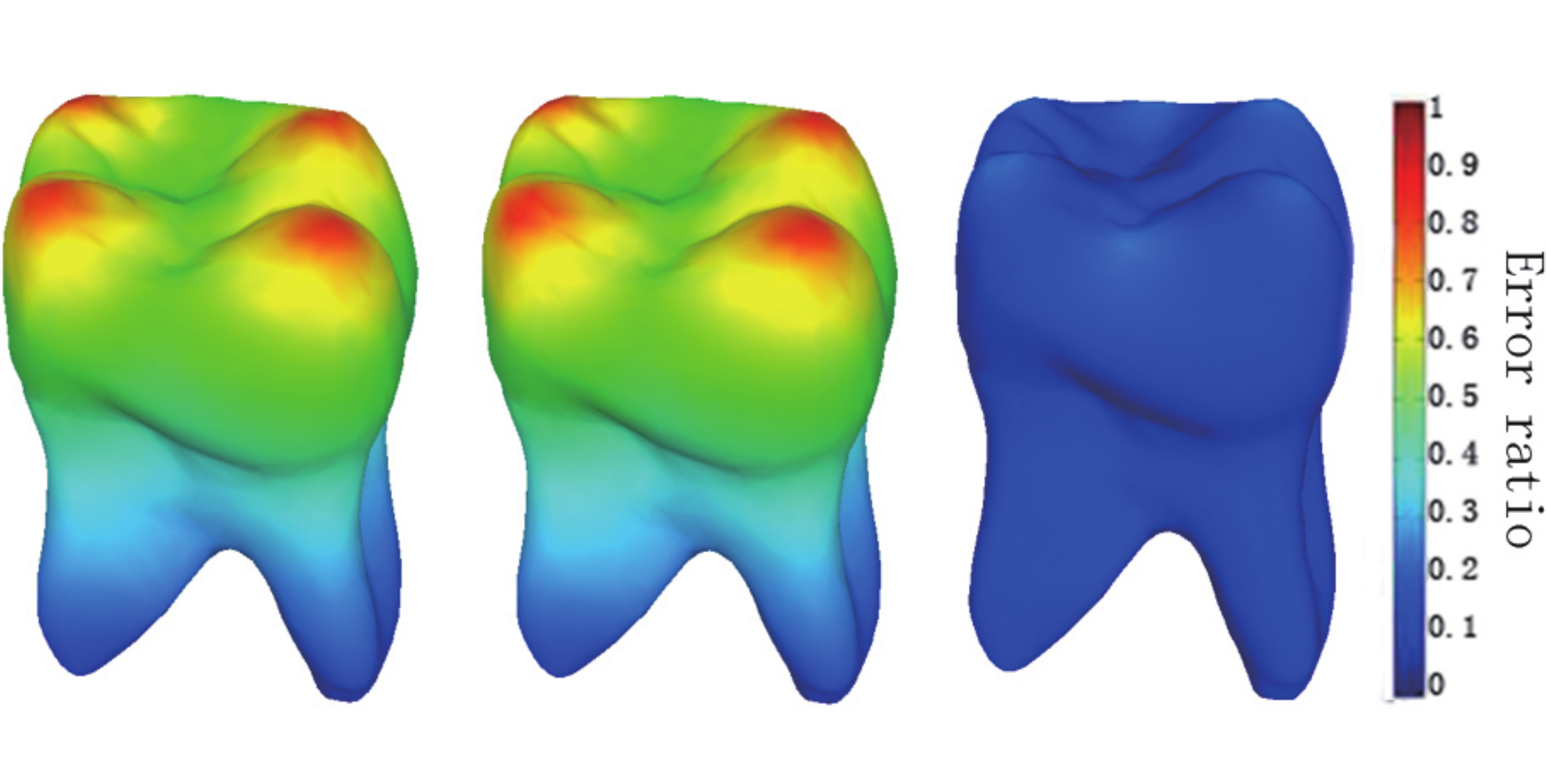}
}
\caption{Visualization results of the distance fields corresponding to tooth landmarks. From left to right in the sub-picture are respectively represented as the predicted distance field, the distance field of GT and the difference between the two distance fields.}
\vspace{-0.3cm}
\label{fig:resuldsitcemap}

\end{figure}

\begin{figure}
    \centering
    \subfigure{
    \includegraphics[width=6.5cm]{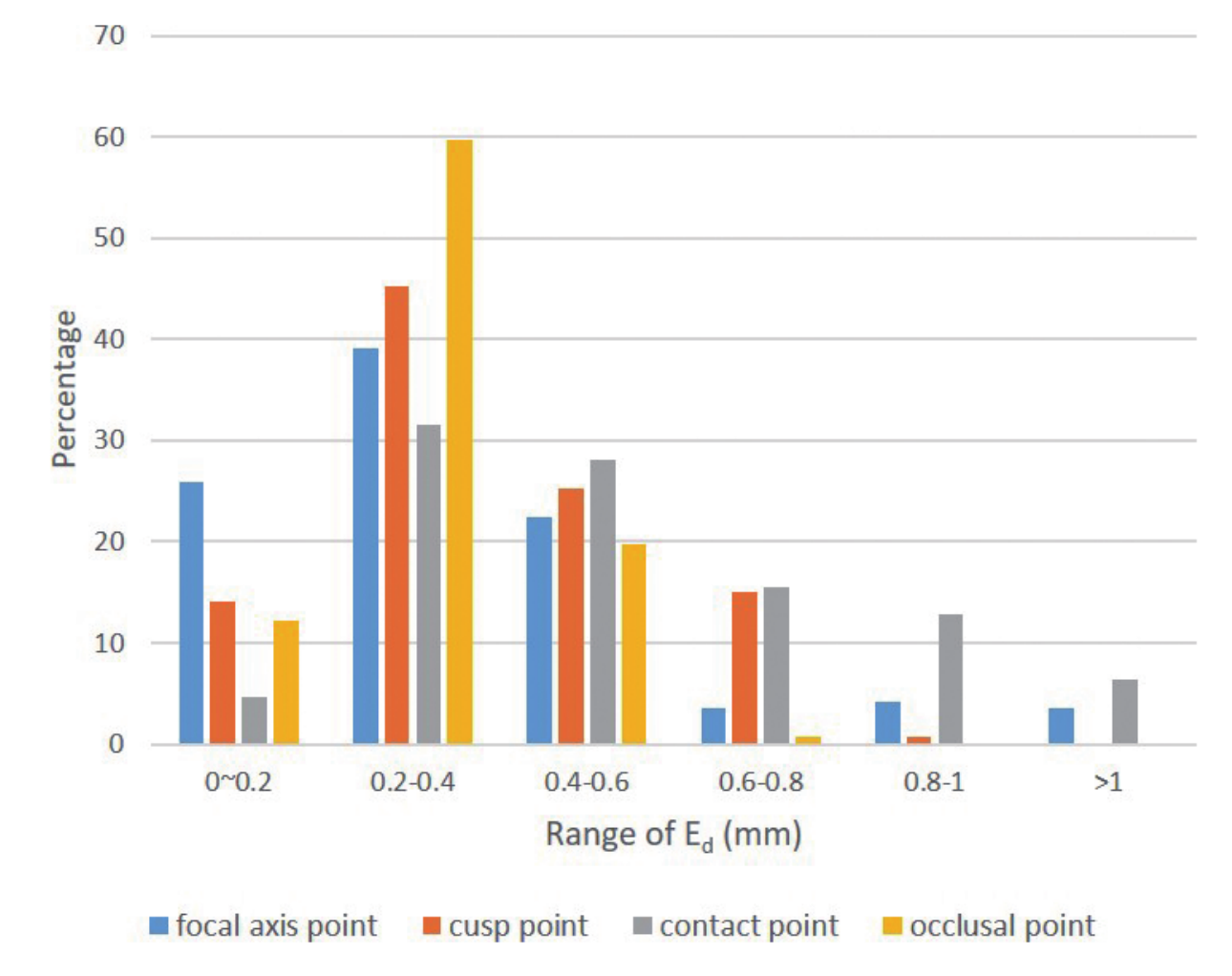}
    }
    \caption{Error probability distribution of test data for landmarks detection.}
    \label{fig:pointhistogram}
    \vspace{-0.3cm}
\end{figure}

\subsection{Dataset.}
To obtain the dataset of 3D tooth models, we first collect 100 CBCT scans and corresponding dental crown models in the real-world clinics and then utilize the data processing to automatically extract the 3D tooth models. The ratio of normal and abnormal occlusion data is approximately 1:1. The dataset contains 3084 3D tooth models, and the corresponding dental landmarks and axes are manually annotated by professional dentists. For each tooth model, we uniformly sample 2048 points on the mesh cells and normalize them within a unit ball. Meanwhile, we calculate the distance field and the projection vectors as the ground truth of our network. In our experiment, we randomly split the data into three subsets with 1855 tooth models for training, 364 tooth models for validation and the rest 865 tooth models for testing.

\begin{figure*}[h]
\centering
\includegraphics[width=6.5in]{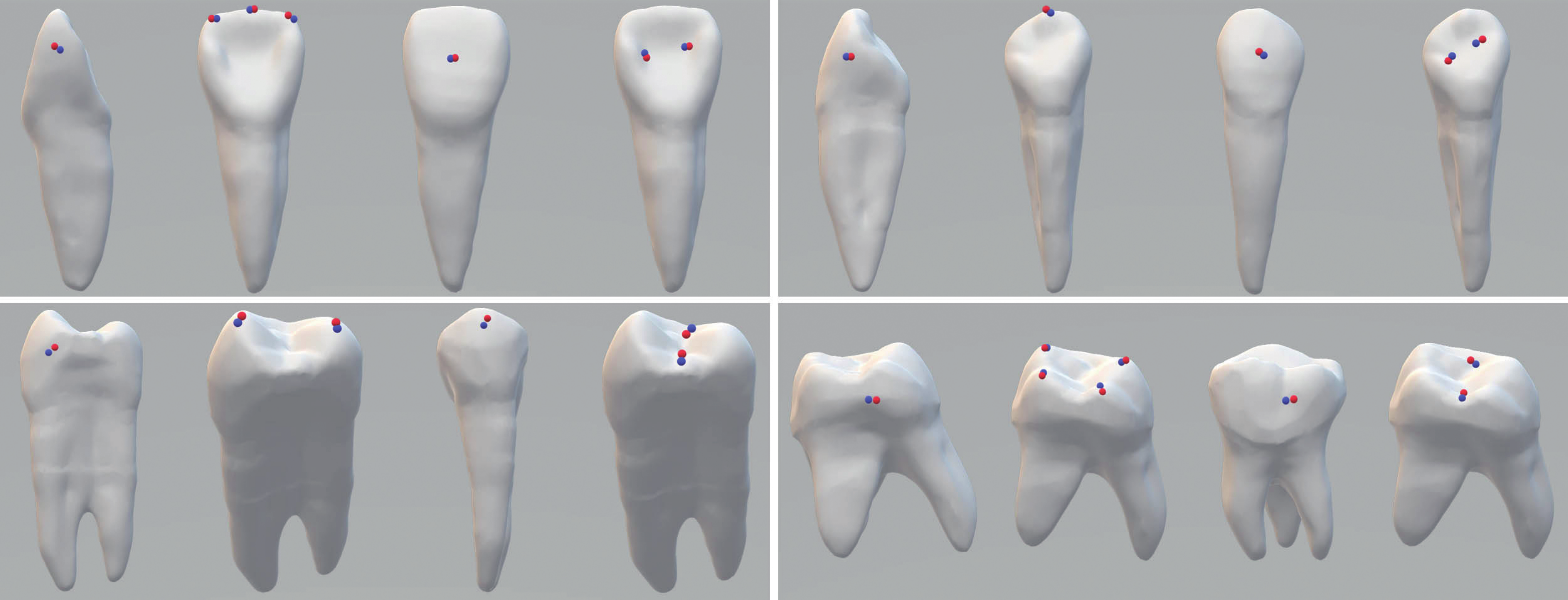}
\caption{The results of the dental landmarks predicted by tooth feature extraction network, where red dots represent GT, blue dots indicate predicted results. We select four representative types of teeth, which are comprised of incisor, canine, premolar, and molar, respectively. Furthermore, each subfigure shows contact point, cusp point, facial axis point, and occlusal point, respectively.}
\label{fig:resultandgt}
%\vspace{-0.2cm}
\end{figure*}

\subsection{Landmarks detection}
\subsubsection{Evaluation metrics}
To evaluate the performance of tooth landmark detection, we employ the average distance error and success rate as the two important metrics to measure the landmark detection accuracy.

\begin{equation}
\label{newerrorpoint}
    E_{d} =\lVert f_{pi} - f'_{pi} \rVert,
\end{equation}
where the $\lVert \cdot \rVert$ represents the Euclidean distance between $f_{pi}$ and $f'_{pi}$. Actually, our network predicts a heatmap for each type of landmark, and the final landmarks are generated by a post-processing step. If there are more than one landmark f one type, for each GT landmark, we define its closest one as the corresponding predicted landmark. We set the parameter $r (mm)$ as the threshold of the average Euclidean distance error, and if $E_{d}$ of a testing tooth is not greater than $r$, it will be considered as a successful case. Thus, the success rate ($SR_p$) is defined as:

\begin{equation}
SR_p=\frac{1}{M_p}\sum_{i=1}^{M_p}\delta_{success}^{i} \times 100\%, \quad \delta_{success}^{i}=
\left\{\begin{matrix}
 1 & \mathrm{if} E_{d} \leq r \\
 0 & \mathrm{if} E_{d} > r.
\end{matrix}\right.
\end{equation}
where $M_p$ is the number of dental landmarks of the testing dataset.

\subsubsection{Results}
For the task of tooth landmarks detection, our method achieves promising results on landmarks detection tasks, and it can be completely applied to the orthodontic process. In order to verify the accuracy of the distance field prediction, it can be seen from the Fig.~\ref{fig:resuldsitcemap} that there is a relatively small error between the predicted distance field and ground truth. Meanwhile, from the perspective of color distribution, the peaks of the distance field are at the same position, then the position of the final landmark detected by post-processing would match the ground truth perfectly. The average distance error ($E_d$) for tooth landmarks by our method is 0.37 $mm$, which is acceptable by the American Board of Orthodontics (ABO)~\cite{morris2019accuracy}.
In addition, it can be seen that the errors of incisor and canine are a little higher than those of premolar and molar, which is due to shape variations of incisor and canine, especially the crown area, are much larger than the molar teeth. In general, our method can accurately localize the landmarks of different types of teeth.

We also visualize the dental landmark detection results using a probability distribution histogram. Fig.~\ref{fig:pointhistogram} shows the distribution of landmarks error probability, and the ordinate is the percentage of the data falling within this range. It can be found that the facial axis point and the contact point have high errors when the $E_{degree}>0.6 mm$. The reason is that, compared to the cusp point and the occlusal point with clear geometric characteristics, the contact point has unclear geometric information and is greatly affected by adjacent teeth. Similarly, the tooth point is mainly located in the center of the tooth surface (outside the gums). 
Fig.~\ref{fig:resultandgt} shows several representative examples of tooth landmark detection results. It can be seen that our method can accurately detect the tooth landmarks compared to the ground truth.

\subsubsection{Comparisons}
To verify the effectiveness of our method, we compare its performance with some well-known state-of-the-art methods, including those designed for this specific task and other related works on general tasks.

\begin{figure}[h]
\centering
\includegraphics[width=0.9\linewidth]{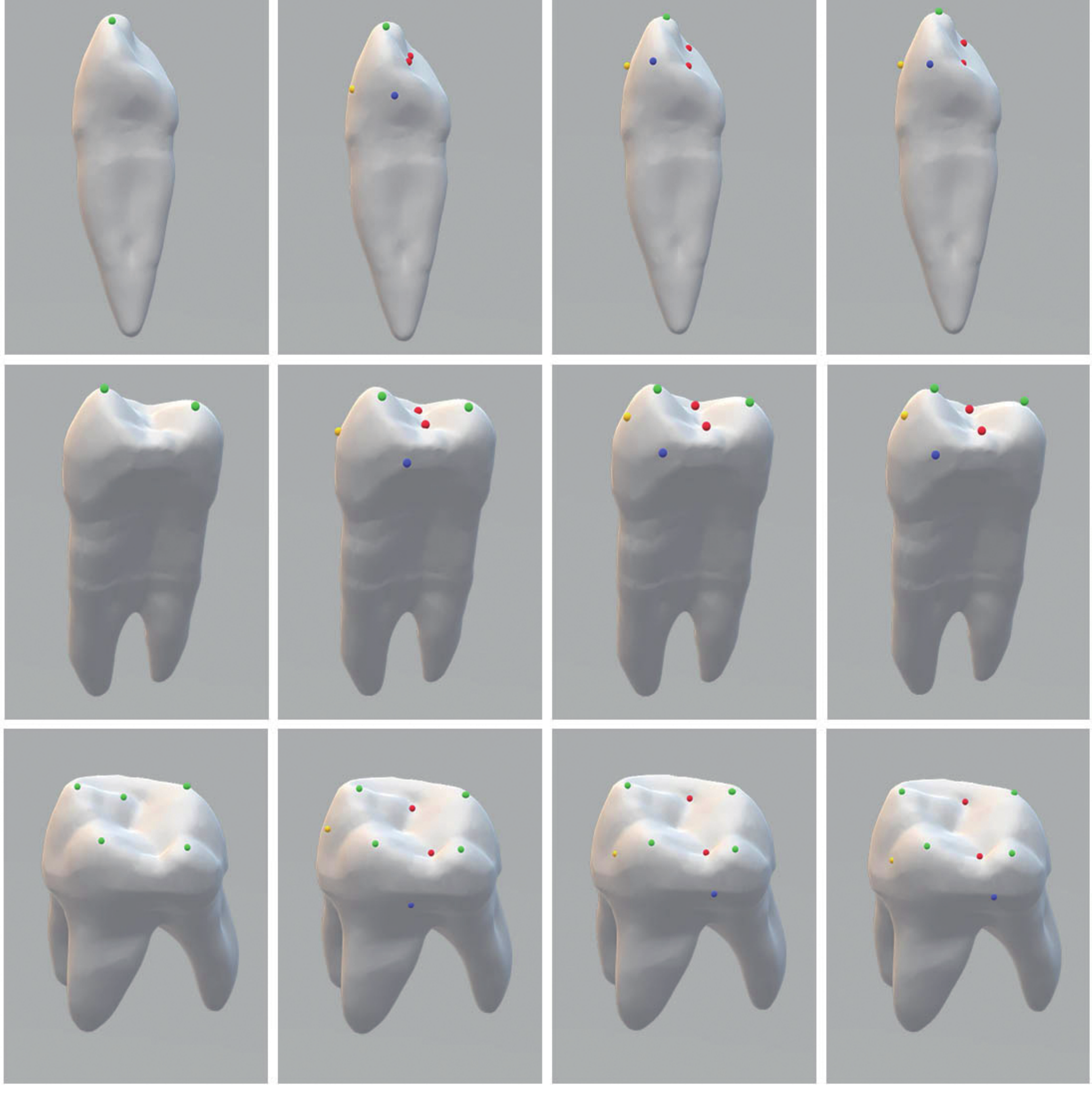}
\caption{A visual comparison of landmarks detection with Kumar \etal~\cite{kumar2012automatic} method (the first column), Shu \etal~\cite{Zhang2016Efficient} method (the second column), our method (the third column) and GT (the fourth column), where the green dots are a cusp point, the blue dots are a contact point, the red dots are an occlusal point and the yellow dots are a facial axis point.}
\label{fig:cusppoint}
\vspace{-0.3cm}
\end{figure}

\noindent\textbf{Comparison with methods developed for dental landmark detection.}
We compare our method with the landmark detection algorithm proposed by Kumar \etal~\cite{kumar2012automatic}, a traditional method that can only detect the landmarks based on the local geometric context, such as cusp point. 
Additionally, we also provide visual comparisons, as shown in Fig.~\ref{fig:cusppoint} and Table~\ref{tab:comparepointothernetwork}, which are consistent with the statistic results and demonstrate the effectiveness of our dense landmark representation.

\noindent\textbf{Comparison with general landmark detection methods.} To further verify the effectiveness of point-wise field coding for 3D landmark detection, we combine our proposed point-wise field coding representation with other network models and use them to predict landmarks of the tooth model. Specifically, we replace the labels of the segmentation parts of other network backbones, such as PointConv~\cite{wu2019pointconv} and SpiderCNN~\cite{xu2018spidercnn}. Besides, we also analyze the performance of the general 3D model feature detection method~\cite{shu2018detecting} on this specific task. 
From Fig.~\ref{fig:cusppoint}, we can see that Shu's method achieves better performance in detecting the obvious position of the cusp point and occlusal point. However, for other areas with weak features, the error is relatively large. On the contrary, our method is not only robust to areas with strong features but also has good detection accuracy for weak features.

From Table~\ref{tab:comparepointothernetwork}, which shows the $SR_p$ in different distance error ranges, we can see that it is often unsatisfactory to directly predict landmark results using only a point cloud-based deep neural network framework.

\begin{table}
\begin{center}
\caption{ The comparison of different networks for tooth landmarks detection based on the $SR_p$ (\%), where $r$ is set to 0.2 $mm$ -1 $mm$. }
\begin{tabular}{lllllllllll}
\hline
 & 0.2 & 0.4 &  0.6 &  0.8  & 1 \\
\hline
PointNet++ &	1.5 &		6.4 &	13.8 &		24.9 &	31.9 \\
PointConv & 	0.9 &		6.9 &		14.5 &	25.8 &		32.7 \\
SpiderCNN & 	1.5 &		8.3 &		16.9 &		29.9 &	36.5 \\
Kumar \etal (cusp points) & 	9.7 &		36.4 &		53.7 &	68.8 &		81.3 \\
Shu \etal & 	3.5 &	19.7 &		38.9 &	52.3 &	67.7 \\
Our method &   \textbf{16.8} & 	\textbf{65.7} &	\textbf{86.3} &	\textbf{94.3} &	\textbf{97.4} \\
\hline
\end{tabular}
\label{tab:comparepointothernetwork}
\end{center}
\end{table}

\subsection{Tooth axes detection}
\subsubsection{Evaluation metrics}
To measure the accuracy of tooth axis detection, we define the axis error $E_{degree}$ as the \emph{arccos} value of the predicted tooth axis $\mathbf{L}_{pi}$ and the ground truth axis $\mathbf{L}_{gi}$.

\begin{equation}
\label{newerroraxis}
E_{degree} =  \frac{1}{M_a}\sum_{i=1}^{M_a} arccos\left( \mathbf{L}_{pi} \cdot \mathbf{L}_{gi} \right).    
\end{equation}
\noindent where $M_a$ is the number of tooth axes of test dataset. 
We set the parameter $dg$ as the threshold of the average axis error, and if $E_{degree}$ of a testing tooth is not greater than $dg$, it will be considered as a successful case. Thus, the success rate (SR\_a) is defined as:

\begin{equation}
\begin{aligned}
SR_a=\frac{1}{M_a}\sum_{i=1}^{M_a}\delta_{success}^{i} \times 100\%, \\ 
\text{where}\quad \delta_{success}^{i}=
\left\{\begin{matrix}
 1 & \mathrm{if} E_{degree} \leq dg \\
 0 & \mathrm{if} E_{degree} > dg.
\end{matrix}\right.
\end{aligned}
\end{equation}

\subsubsection{Results}
\begin{figure}[h]
\centering
\includegraphics[width=0.8\linewidth]{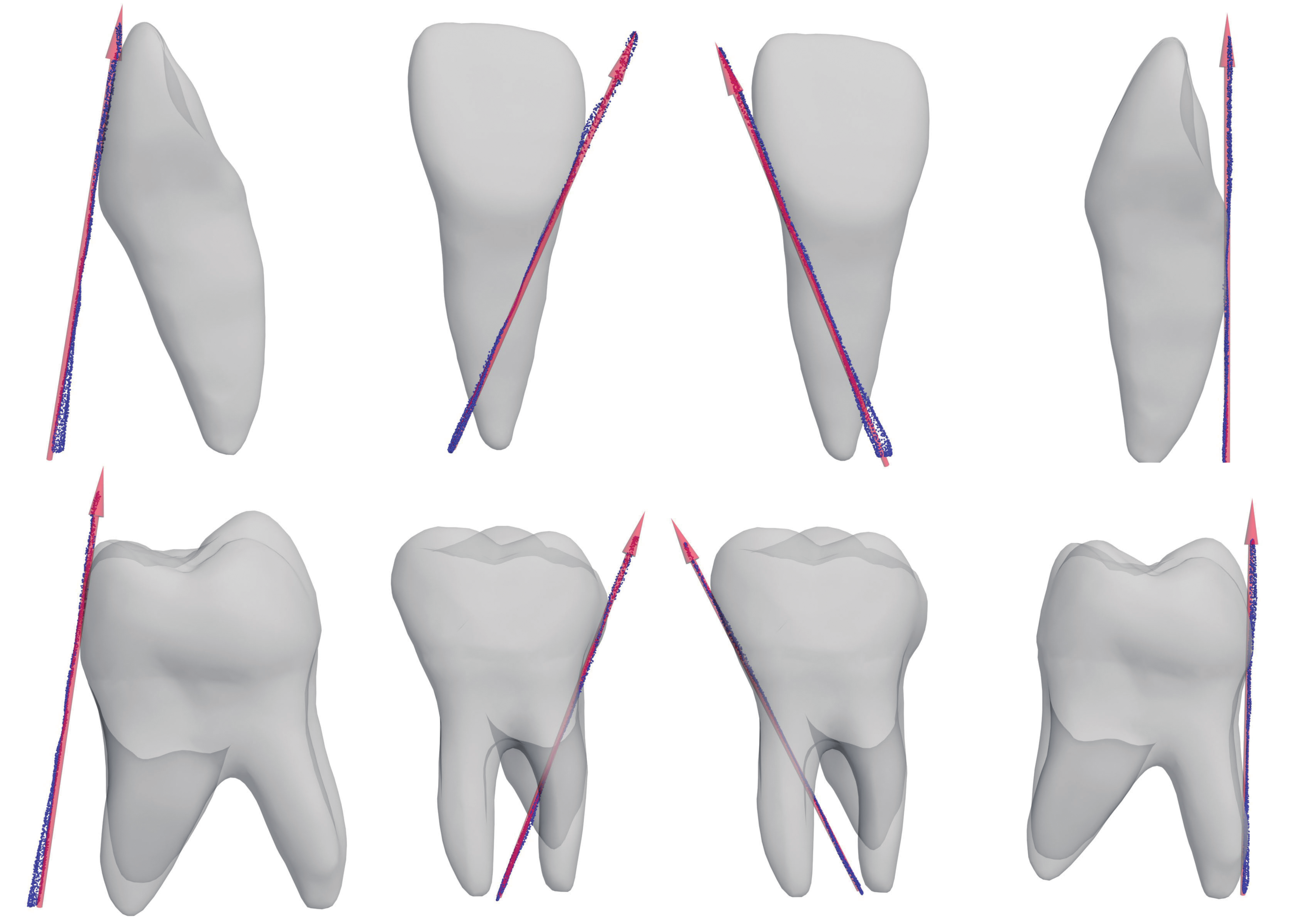}
\caption{Visualization results of the projection vectors corresponding to tooth axes, where the blue dotted lines are the predicted projection vectors and the red dotted lines are GT. Each rows shows buccal surface axis (BA), lingual axis (LA), mesial axis (MA) and distal axis (DA), respectively.}
\label{fig:resultproline}
%\vspace{-0.2cm}
\end{figure}

For tooth axes detection, we design another branch network to predict projection vectors. It can be seen from Fig.~\ref{fig:resultproline} that the displacement point cloud set obtained by point-wise projection vectors, and the direction is similar to the ground truth.

\begin{figure}[h]
    \centering
    \includegraphics[width=6.5cm]{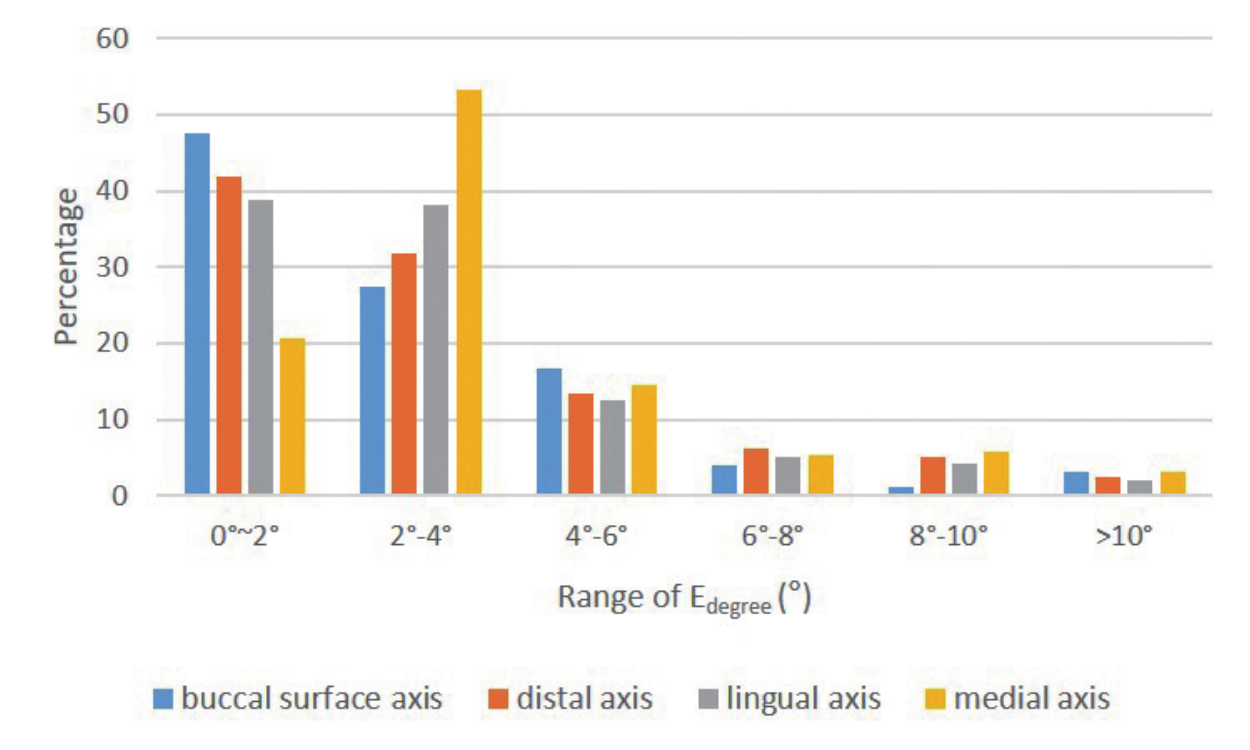}
    \caption{Error probability distribution of test data for tooth axes detection.}
    \label{fig:histogram}
    \vspace{-0.3cm}
\end{figure}

Fig.~\ref{fig:histogram} shows the tooth axes error probability, where the range of the abscissa is set to 2\degree. From Fig.~\ref{fig:histogram}, it can be found that there is no large fluctuation in the results of tooth axes detection, and the errors are concentrated in a small range.

\begin{figure*}[h]
\centering
\includegraphics[width=0.9\linewidth]{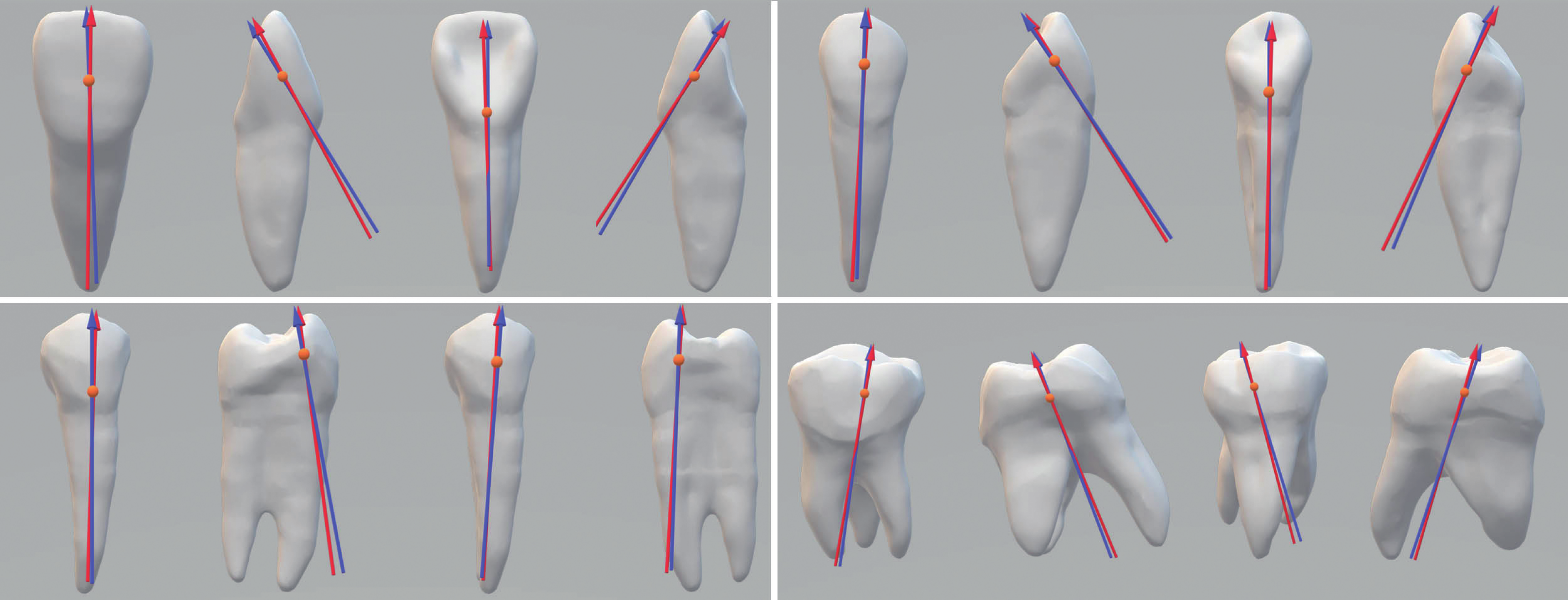}
\caption{The results of the tooth axes predicted by our method, where red arrows represent GT, blue arrows indicate predicted results. We select four representative types of teeth, which are comprised of incisor, canine, premolar and molar, respectively. Furthermore, each sub figure shows buccal surface axis, lingual
axis, mesial axis and distal axis, respectively.}
\label{fig:resultandgtaxes}
%\vspace{-0.2cm}
\end{figure*}

We also visualize the predicted tooth axes results from the test dataset as shown in Fig.~\ref{fig:resultandgtaxes}. It can be seen that the tooth axes are mainly dependent on the global tooth context instead of the local information.
And compared with the ground truth annotated by the dentist, our method can still accurately detect the tooth axes.

\begin{table}[h]
\begin{center}
\caption{ Comparison of different networks for tooth axes detection based on $E_{degree}$. }
\begin{tabular}{llllll}
\hline
 & BA & MA & DA & LA & Average\\
\hline
PointNet++ & 37.45& 52.91 & 47.55 & 48.46 & 46.59\\
PointConv & 30.57 & 38.28 & 45.77 & 42.19 & 39.19 \\
SpiderCNN & 32.43 & 36.46 & 44.92 & 41.78 & 38.89\\
PCPNet & 11.36 & 12.21 & 10.97 & 11.62 & 11.54\\
Our method & \textbf{3.07} & \textbf{3.53} & \textbf{3.58} & \textbf{3.12} & \textbf{3.33}\\
\hline
\end{tabular}
\vspace{-0.3cm}
\label{tab:compareaxistableothernetwork}
\end{center}
\end{table}

\begin{table}[h]
\begin{center}
\caption{The comparison of different networks for tooth axis detection based on the $SR_a$ (\%), where $dg$ is set to 2\degree-10\degree.}
\begin{tabular}{lllllllllll}
\hline
 &  2\degree &  4\degree &  6\degree &  8\degree &  10\degree \\
\hline
PointNet++ &	5.9 &	9.2 &	15.8 &		20.9 &	28.9 \\
PointConv & 	6.1 &	12.4 &	17.6 &		27.2 &		34.4 \\
SpiderCNN & 5.8 &9.2 &	14.7 &20.6	 & 28.5	 \\
PCPNet & 8.9 & 25.7 &   38.2 &   49.1  &64.4 \\
Our method &  \textbf{38.4} & 	\textbf{78.2} &	\textbf{90.3} &		\textbf{94.2} &		\textbf{96.9} \\
\hline
\end{tabular}
\label{tab:compareaxisothernetwork}
\end{center}
\vspace{-0.3cm}
\end{table}

\subsubsection{Comparisons}
To the best of our knowledge, it is the first time that we use deep learning to detect tooth axes. Therefore, similar to the comparison of landmarks detection, we use several typical point cloud based network frameworks to regress tooth axes and have a comparison. Furthermore, we also compare with a point cloud normal estimation method (PCPNet~\cite{GuerreroKOM18}), which is highly related to our tooth axis detection task. Specifically, we use this method to predict the tooth axis vector of each point instead of the original normal vector. Then, the average of the predicted axis vector at each point is defined as the final tooth axis. From Table~\ref{tab:compareaxistableothernetwork} and Table~\ref{tab:compareaxisothernetwork}, it can be seen that, compared to the point cloud networks that directly regress tooth axes, PCPNet has significantly improved the accuracy demonstrating the effectiveness of dense representation. In addition, our algorithm has achieved the best performance (3.33\degree), indicating that our tooth axis dense representation (projection vector field) is more suitable in this specific task, compared to the straightforward point-wise representation (PCPNet).

\begin{table}
\begin{center}
\caption{ The average distance error of tooth landmarks and axes detection based on $E_d$ and $E_{degree}$. PFC refers to point-wise field coding representation. FA refers to the feature aggregation module, CA is the cross attention mechanism. FE represents the feature enhancement module }
\begin{tabular}{llllll}
\hline
 & FA & OC & CO & CU & Average \\
\hline
w/o PFC & 8.6 & 9.7 & 9.4 & 9.5 & 9.3\\
 w/o FA & 0.78 & 0.81 & 0.95 & 0.58 & 0.78\\
 w/o CA & 0.49 & 0.39 & 0.64 & 0.38 & 0.48\\
 w/o FE & 0.48 & 0.36 & 0.53 & 0.33 & 0.43\\
 Our method & \textbf{0.45} & \textbf{0.24} & \textbf{0.51} & \textbf{0.29} & \textbf{0.37}\\
\hline
 & BA & MA & DA & LA & Average \\
\hline
w/o PFC & 31.25 & 43.14 & 46.58 & 39.72 & 40.17\\
w/o FA & 5.23 & 6.12 & 6.21 & 6.28 & 5.96\\
w/o CA & 3.64 & 4.23 & 4.17 & 4.03 & 4.02\\
w/o FE & 3.26 & 3.77 & 3.91 & 3.42 & 3.59\\
Our method & \textbf{3.07} & \textbf{3.53} & \textbf{3.58} & \textbf{3.12} & \textbf{3.33}\\
\hline
\end{tabular}
\label{tab:Tabresultpoint}
\end{center}
\vspace{-0.3cm}
\end{table}

\subsection{Ablation Study}

To verify the efficiency of the framework with different components, including the multi-scale feature extraction module and feature enhancement module, we validate the effectiveness of the results of the different modules by removing each module separately. All alternative networks are trained on the same dataset and evaluated on the testing dataset for comparison.

\subsubsection{Point-wise field coding representation} 
To validate the importance of point-wise field coding, we first train a network that directly regresses the 3D landmark coordinates and axis vector by the same network backbone. The other way is to encode the dense tooth feature representation into landmarks and axes. In addition, an interesting finding is that the network based on dense feature representation outperforms the method that directly regresses the landmarks by a large margin (9.3 $mm$ vs 0.37 $mm$) in Table~\ref{tab:Tabresultpoint}. It means that the dense outputs provide rich information and reduce the uncertainty compared to the direct regression. In conclusion, quantitative results on landmark and tooth axis detection demonstrate the effectiveness of point-wise field coding and multi-scale feature detection.

\begin{figure*}[h]
\begin{center}
\includegraphics[width=0.9\linewidth]{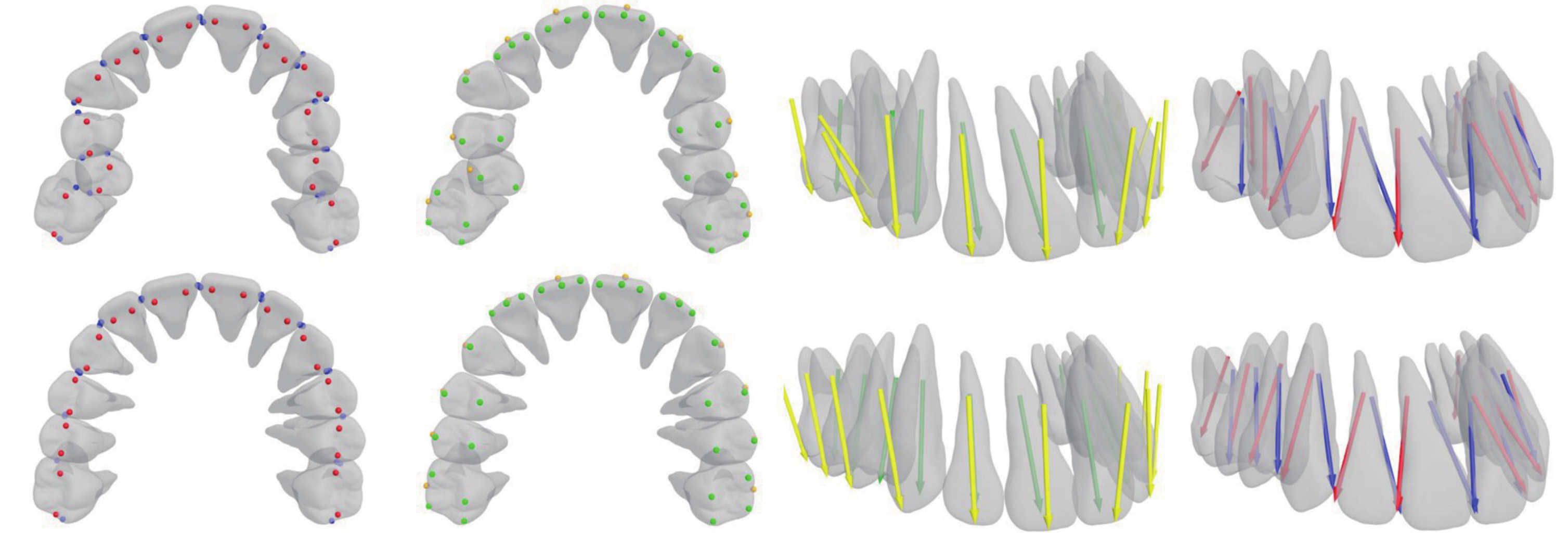}
\end{center}
\caption{Results of the treatment planning before and after orthodontics, where the first row represents the sequence of the teeth before arrangement and the second row indicates the corrected tooth arrangement based on the landmarks and axes detected by our method.}
\label{fig:alltooth}
\vspace{-0.3cm}
\end{figure*}

\subsubsection{Multi-scale feature extraction module} The multi-scale feature extraction module can extract local and non-local features from the tooth model based on a feature aggregation mechanism (FA) and the cross attention module (CA), which can effectively detect landmarks located in regions without obvious geometric features. To verify the performance of those modules, we remove single modules of the full network and keep the other modules unchanged for experimental comparison. From Table~\ref{tab:Tabresultpoint}, it can be found that FA and CA have a greater effect on tooth landmark/axis detection. Especially for the landmarks located in the smooth tooth area, the accuracy is significantly improved. Similarly, as shown in Table~\ref{tab:Tabresultpoint}, it can be found that the accuracy of the tooth axes is also improved by a multi-scale scheme.

\subsubsection{Feature enhancement module} The feature enhancement module mainly uses multiple skipped connections to concatenate the original input information and the feature map. In order to verify the effectiveness of this module on the entire network structure, we compare it with the version by deleting all skipped connections. From Table~\ref{tab:Tabresultpoint}, we can see that the feature enhancement module improves the accuracy of tooth landmarks (average $E_{d}$: 0.43 $mm$ vs. 0.37 $mm$) and axes (average $E_{degree}$: 3.59\degree vs. 3.33\degree).

\subsection{Discussion}
\subsubsection{Verification of landmarks and axes}
To verify the effectiveness of our method for predicting landmarks and axes, we simulate orthodontic treatment based on the rules of Andrew’s six keys of occlusion~\cite{andrews1972six} and the features of the teeth obtained by our method. Firstly, we get a set of tooth models before the orthodontics treatment. Then, we predict the features of teeth by our network and align a set of teeth according to the predicted landmarks and tooth axes, as shown in Fig.~\ref{fig:alltooth}.
It can be seen that the outputs after planning are more reasonable and meaningful as compared to pre-treatment teeth alignment~\cite{dai2018complete}. Therefore, it verifies the importance and practicality of the landmarks and axes detected by our method in the teeth arrangement task, which can free the dentists from labeling the tooth landmarks and axes manually.

To further verify the clinical applicability of our method, we also collect 5 external sets of tooth models, and compared them with the performance of the manual process. Note that the ground truth was annotated by three senior dentists after consultation, and competing them with dentists who did not participate in the labeling process. As shown in Table~\ref{tab:com2dentist}, compared with the manual annotation, the average time consumed is greatly reduced from $\sim$1min to 0.14s. Simultaneously, the accuracy is more stable compared to the ground truth. The industrial and clinical partners confirm that such a performance is incontrovertibly acceptable and convenient for the downstream orthodontic treatment.

\begin{table}[h]
\begin{center}
\caption{  The average error of tooth landmarks/axes detection based on $E_d$/$E_{degree}$. $T(S)$ represents the time (second) to calculate/mark a tooth landmark point and axis.   }
\begin{tabular}{llllll}
\hline 
 & dentist1 & dentist2 & dentist3 & Our method \\
\hline
Landmark(mm) & 0.64 & 0.52 & 0.55 & 0.43  \\
$T(s)$ & 32.86 & 43.14 & 35.53 & 0.14 \\
Axis(\degree) & 5.47 & 4.34  & 4.66 & 3.51 \\
$T(s)$ & 51.47 & 86.22 & 74.37 & 0.14 \\
\hline
\end{tabular}
\vspace{-0.3cm}
\label{tab:com2dentist}
\end{center}
\end{table}

\subsubsection{Limitations}
During the network training process, our input data are the point cloud of a single complete tooth extracted from the CBCT. However, if the tooth model is partially damaged, the predicted result will be worse. So our method highly relies on the quality of the input tooth data.

\section{Conclusion}
This is the first work to utilize deep neural networks to solve the problem of landmark and axis detection on 3D tooth model, which plays an irreplaceable role in orthodontics. Aiming at the challenge of sparse features learning, we design a point-wise field coding method to overcome the difficulty of predicting sparse features based on deep learning cleverly. Our method enables the automatic detection of landmarks and axes without manual intervention. We also adopt multi-scale feature extraction and feature enhancement modules to faithfully learn the local and global features. Extensive experimental results show that our method achieves state-of-the-art performance, which gives the potential to be used in real-world clinics.

\bibliographystyle{dentalfeature}
\bibliography{dentalfeature}

\end{document}